\documentclass[sigconf]{acmart}

\usepackage{multirow}
\usepackage{makecell}
\usepackage[nameinlink]{cleveref}
\usepackage{xspace}
\usepackage{soul}
\usepackage{wrapfig}

\newcommand{\systembf}{\textbf{S\&UI}\xspace}
\newcommand{\system}{S\&UI\xspace}

\AtBeginDocument{%
  }

\copyrightyear{2025}
\acmYear{2025}
\setcopyright{cc}
\setcctype{by}

\acmConference[CHI '25]{CHI Conference on Human Factors in Computing
Systems}{April 26-May 1, 2025}{Yokohama, Japan}

\acmBooktitle{CHI Conference on Human Factors in Computing Systems (CHI '25), April 26-May 1, 2025, Yokohama, Japan}
\acmDOI{10.1145/3706598.3714213}
\acmISBN{979-8-4007-1394-1/25/04}

\newcommand{\eg}{\textit{e.g.}\xspace}
\newcommand{\ie}{\textit{i.e.}\xspace}

\newcommand{\etal}{\textit{et al.}\xspace}

\definecolor{tableheader}{HTML}{EFEFEF}
\definecolor{tablegrayline}{HTML}{d0d0d0}

\definecolor{darkgray}{HTML}{555555}

\definecolor{diffbg}{RGB}{242, 196, 123}
\definecolor{difftxt}{RGB}{127, 17, 16}
\definecolor{nondiffbg}{RGB}{255, 255, 255}
\definecolor{nondifftxt}{RGB}{0,0,0}

\begin{document}

\title[Leveraging Multimodal LLM for Inspirational User Interface Search]{Leveraging Multimodal LLM \\for Inspirational User Interface Search}

\author{Seokhyeon Park}
\orcid{0009-0003-1685-4027}
\affiliation{%
  \institution{Seoul National University}
  \city{Seoul}
  \country{Republic of Korea}
}
\email{shpark@hcil.snu.ac.kr}

\author{Yumin Song}
\orcid{0009-0004-5277-4822}
\affiliation{%
  \institution{Seoul National University}
  \city{Seoul}
  \country{Republic of Korea}
}
\email{ymsong@hcil.snu.ac.kr}

\author{Soohyun Lee}
\orcid{0000-0002-3075-3981}
\affiliation{%
  \institution{Seoul National University}
  \city{Seoul}
  \country{Republic of Korea}
}
\email{shlee@hcil.snu.ac.kr}

\author{Jaeyoung Kim}
\orcid{0009-0006-1868-7148}
\affiliation{%
  \institution{Seoul National University}
  \city{Seoul}
  \country{Republic of Korea}
}
\email{jykim@hcil.snu.ac.kr}

\author{Jinwook Seo}
\authornote{Corresponding Author}
\orcid{0000-0002-7734-822X}
\affiliation{%
  \institution{Seoul National University}
  \city{Seoul}
  \country{Republic of Korea}
}
\email{jseo@snu.ac.kr}

\begin{abstract}
Inspirational search, the process of exploring designs to inform and inspire new creative work, is pivotal in mobile user interface (UI) design.
However, exploring the vast space of UI references remains a challenge.
Existing AI-based UI search methods often miss crucial semantics like target users or the mood of apps.
Additionally, these models typically require metadata like view hierarchies, limiting their practical use.
We used a multimodal large language model (MLLM) to extract and interpret semantics from mobile UI images.
We identified key UI semantics through a formative study and developed a semantic-based UI search system.
Through computational and human evaluations, we demonstrate that our approach significantly outperforms existing UI retrieval methods, offering UI designers a more enriched and contextually relevant search experience.
We enhance the understanding of mobile UI design semantics and highlight MLLMs' potential in inspirational search, providing a rich dataset of UI semantics for future studies.

\end{abstract}

\ccsdesc[500]{Human-centered computing~User interface design}
\ccsdesc[500]{Human-centered computing~Interactive systems and tools}
\ccsdesc[500]{Information systems~Multimedia and multimodal retrieval}

\keywords{Interface Design, UI Design, UI Retrieval, UI Search, Semantic Search, Multimodal LLM}

\begin{teaserfigure}
\centering
  \includegraphics[width=\textwidth]{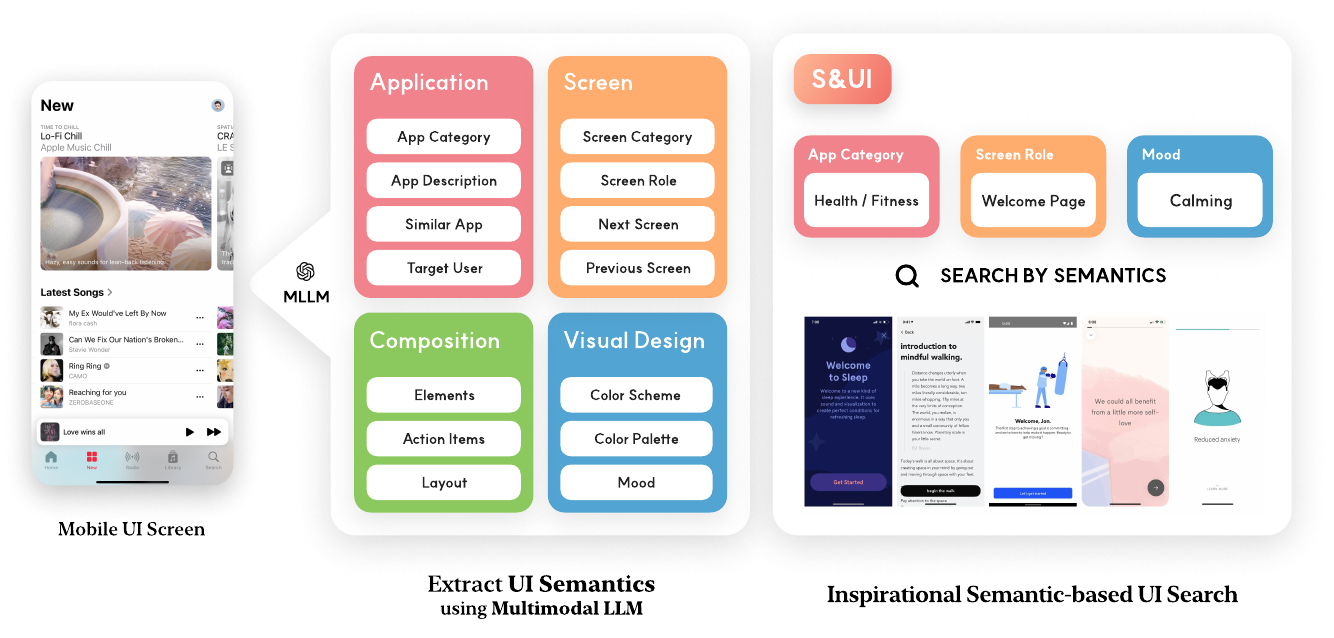}
  \caption{Overview of our approach for semantic-based UI search. Left: A multimodal large language model is employed to extract UI semantics from mobile UI screens. Right: S\&UI, an inspirational semantic-based UI search system, enables designers to find relevant UI designs by specifying desired semantic attributes.}
  \Description{Overview of the semantic-based user interface (UI) search approach. On the left, a multimodal large language model (MLLM) extracts semantic attributes from mobile UI screens. On the right, an inspirational semantic-based UI search interface allows designers to find relevant UI designs by specifying desired semantic attributes, such as App Category, Screen Role, and Mood.}
  \label{fig:teaser}
\end{teaserfigure}

\maketitle

\section{Introduction}

User interface (UI) design plays a central role in creating intuitive and engaging digital experiences.
A vital part of the UI design process is the inspirational search, where designers explore existing UI designs to inform and inspire their own work~\cite{herring2009getting, siangliulue2015providing}.
By leveraging a wide array of reference materials, designers can craft innovative, user-centered interfaces that build upon established design patterns and best practices~\cite{chan2018best, gonccalves2014inspires, mayai, bonnardel1999creativity}.
During inspirational search process for UI design, designers often focus on functionalities, problem domains, and visual styles~\cite{lu2022bridging}.
These aspects offer valuable insights into the design choices of UIs and assist designers in comprehending how to fulfill user needs and expectations effectively.

However, current UI retrieval methods are limited in their ability to capture the nuanced semantic elements essential to an effective, inspirational design process.
Prior research~\cite {SWIRE, guing} have focused on pixel-level retrieval methods, emphasizing visual similarities.
While these approaches are effective for identifying surface-level design patterns, they often overlook in-depth semantic aspects such as user interaction flow, target audience, and the emotional tone of an application.
Some computational methods~\cite{RICO-semantic, Screen2Vec, VINS} have attempted to address these limitations by incorporating metadata such as view hierarchies and text labels to enable more comprehensive browsing of UI semantics.
However, the effectiveness of these methods is constrained by the quality and availability of metadata, which can vary widely across different UI datasets and design tools, limiting their scalability and applicability.
Likewise, platforms like Pinterest, Behance, and Dribbble are popular due to their vast collections of design examples and are frequently used for visual inspiration.
However, these platforms primarily offer image-based content with limited ability to filter by deeper semantic attributes, making it challenging for designers to find contextually relevant UI references~\cite{wu2021exploring}.
UI curation platforms like Mobbin, which categorize designs by app information, screen category, and UI elements, offer some improvements but rely heavily on manual curation and human annotation, providing limited scalability. %

To address the limitations of existing UI search methods, we propose a novel approach that leverages a multimodal large language model (MLLM) to extract semantic information directly from UI images without relying on additional metadata.
MLLMs are powerful models that can process and integrate information from multiple modalities, such as images and text~\cite{wu2023multimodal, yin2023survey}.
By leveraging the visual and textual features presented in UI images, MLLMs can infer complex semantics and generate meaningful metadata without manual annotation.
By focusing on images without any additional information, this approach significantly improves scalability, allowing for seamless application across diverse UI datasets and design tools.
Given the potential of MLLMs to offer a richer and more comprehensive understanding of design elements, this approach provides a more scalable, flexible, and efficient solution compared to traditional methods that rely on metadata.
This approach enables the generation of new, structured metadata from raw UI images, expanding the possibilities for other types of UI understanding.

To explore the potential of this approach, we begin our research by investigating the following key research questions:

\begin{enumerate}
\raggedright
\item[\textbf{RQ1}] What UI semantics are most important for UI designers during their search for inspiration?
\item[\textbf{RQ2}] How accurately can a multimodal large language model (MLLM) extract semantic elements from UI images?
\item[\textbf{RQ3}] What are the advantages of semantic-based UI search compared to traditional pixel or keyword-based retrieval?
\end{enumerate}

To answer these research questions, we first conducted a formative study with designers to identify the most crucial semantic facets of inspirational search.
Based on these findings, we developed a novel methodology for extracting and structuring semantic information using MLLM.
Utilizing these extracted semantics, we designed \system, a semantic-based UI retrieval system.
To assess the effectiveness of our approach, we conducted a multi-stage evaluation process.
This evaluation included computational assessments using established UI datasets to measure the accuracy of app and screen category classification, UI element prediction, and screen role understanding.
We also performed human evaluations with designers, combining quantitative ratings and qualitative insights to validate the quality and usefulness of the extracted semantics.
Furthermore, we compared our semantic-based UI search method to existing pixel and keyword-based retrieval methods.
Through this comparative study, we demonstrated the advantages of our approach across multiple aspects, showcasing its effectiveness in inspirational UI search through quantitative ratings and qualitative insights from designer interviews.

Building on these efforts, we highlight three key contributions:
\begin{itemize}
    \item We identify the key UI semantics that designers prioritize during UI inspirational search through a formative study.
    \item We propose a novel MLLM-based approach for extracting rich semantics directly from UI screen images and assess its effectiveness through computational and human evaluations.
    \item We introduce \system, a semantic-based UI retrieval system, and demonstrate its significant strengths over existing pixel and keyword-based UI search methods in multiple criteria.
\end{itemize}

\section{Related Work}

\subsection{Inspirational Search in UI Design Process}
The user interface design process is a complex and iterative endeavor that involves multiple stages, from ideation to implementation~\cite{Rogers2023InteractionDB}.
This complexity drives designers to search for external sources of inspiration that can inform their design decisions and stimulate creativity~\cite{chan2018best, gonccalves2014inspires}.
Herring \etal~\cite{herring2009getting} highlighted the importance of examples as a source of inspiration in the design process, and multiple studies~\cite{bonnardel1999creativity, siangliulue2015providing, lee2010designing} noted their role in enhancing designers' creativity and leading to more diverse and innovative solutions.

During the inspirational search for UI design, designers need to understand key semantic aspects, including functionalities, problem domains, and visual styles~\cite{lu2022bridging}.
However, current platforms struggle to provide these aspects.
Wu and Xu \etal~\cite{wu2021exploring} noted that current creative platforms such as Behance and Dribbble provide vast collections focusing primarily on visual style inspiration but lacking UX-oriented examples. On the other hand, UI-specific platforms like Mobbin lack options for visual style exploration and attempt to address functionality by manually curating and annotating designs, which inherently limits their scalability and comprehensiveness.
Due to these limitations, designers struggle to find examples that meet their semantic requirements, which is particularly critical in fast-paced design environments~\cite{ritchie2011d, sharmin2009understanding}.
Therefore, there is a need for a new approach that identifies those semantics and supports rich and extensible inspiration searches.

To address this need, this study explores the application of MLLMs in UI inspiration search, leveraging their visual understanding and contextual reasoning capabilities.
By supporting semantic analysis of UI elements—such as interpreting visual styles, domain, and functional roles—this approach aims to enable a multidimensional inspiration search, addressing gaps in existing methods.

\subsection{Computational UI Retrieval}
The field of computational UI understanding has seen significant advancements in recent years, driven by the increasing availability of UI datasets (\eg, Rico~\cite{RICO, RICO-semantic} and its curated subset Enrico~\cite{enrico}) and the development of various deep learning techniques~\cite{computational_ui_workshop}.
One of the key tasks within this field is UI retrieval, which focuses on efficiently identifying and recovering relevant user interfaces.

Several works have focused on learning latent semantic embeddings for UI components and screens, enabling more effective retrieval and clustering of similar UIs.
Deka \etal~\cite{RICO-semantic} introduced a method for learning semantically meaningful embeddings of UI components, allowing for efficient search and analysis of UI design patterns.
Bunian \etal~\cite{VINS} introduced VINS, a UI design search system that utilizes wireframes or screenshots to query and retrieve relevant UI examples.
Huang \etal~\cite{SWIRE} proposed Swire, a similar system that combines sketch-based retrieval with UI component recognition to enable more flexible and intuitive search experiences.
Methods like Screen2Vec~\cite{Screen2Vec} generate embeddings that integrate UI components, layouts, and texts for robust retrieval. 
These approaches enabled retrieval of semantically similar screens by training latent representation of semantics.
However, they are utilizing not only UI images but also metadata, such as view hierarchies, which are often noisy~\cite{CLAY} and inherently limit their scalability.
Recognizing this challenge, recent machine learning approaches have shown another direction.
Park and Kim \etal~\cite{spark2023ICML} demonstrated that large-scale foundation models like CLIP~\cite{CLIP}, trained without relying on additional UI-specific metadata, can achieve superior retrieval performance compared to traditional approaches.
As a stretching of this approach, GUIClip~\cite{guing}, a CLIP model fine-tuned with UI datasets, also showed better retrieval performance than prior methods.

However, most of the above approaches rely on embeddings, which often fail to fully reflect human perception~\cite{soohyun}, and their focus on embedding similarity limits both interpretability and depth of semantic understanding.
Building on the strengths of foundation models, the emergence of MLLMs has opened new possibilities.
Unlike previous methods that utilize latent semantic representations and their similarities, MLLMs can directly generate semantic outputs from UI screenshots, enabling granular and explainable retrieval.
Our work builds upon these foundations while addressing key limitations: reliance on additional annotations and restricted semantic explanation.
So, we investigate whether MLLMs can extract rich semantic information directly from UI screenshots, enabling UI design inspirational search that considers both functional context and design intent beyond surface-level visual patterns.

\subsection{Multimodal Learning and Language Models in UI Design} %

The application of language models and multimodal learning techniques in UI design has gained increasing attention in recent years.
While earlier works primarily focused on visual understanding and retrieval using deep learning methods, the emergence of large language models (LLMs) and vision-language models (VLMs) has opened up new possibilities for assisting various aspects of the UI design process.

Several studies have explored the use of multimodal learning for UI-related tasks, such as generating natural language descriptions and enabling conversational interactions.
Li \etal~\cite{WidgetCaptioning} developed a method for generating natural language descriptions of UI elements, which can aid in accessibility and user understanding.
Prior works like Screen2Words~\cite{Screen2Words} and UIBert~\cite{UIBert} learned embeddings based on the screen images, view hierarchies, and texts in the UI, enabling the generation of natural language descriptions and the discovery of semantic relationships.
These approaches enabled a textual understanding of UI by offering descriptions.
Recently, Wang \etal~\cite{wang2023enabling} demonstrated the potential of LLMs for enabling conversational interactions with mobile UIs, paving the way for users to accomplish tasks through natural language dialogue.

More recent studies have begun to explore the application of MLLMs on UI that go beyond the simple vision-language model.
For example, Ferret-UI~\cite{FerretUI} combines a customized visual sampler with MLLM, enabling UI-related tasks such as referring, grounding, and reasoning of user interfaces.
Likewise, AppAgent~\cite{AppAgent} leverages an MLLM to operate the user interface in a manner similar to human users, enabling autonomous task execution.
While these approaches demonstrate the potential of MLLMs in UI understanding, they are not specifically tailored for UI design tasks.
Furthermore, most of the above approaches require not only images but additional UI metadata (\eg, types of UI elements or coordinates) at the time of inference.
This reliance on metadata makes it impractical to search for UI design inspiration, as the vast majority of inspiration materials lack structured UI metadata.

This study addresses the challenge of effectively capturing and leveraging the semantic elements crucial for designers during inspirational searches and creative processes, using only screen images.
We propose a novel MLLM-based approach for semantic extraction and retrieval, enabling designers to capture the nuanced intentions underlying their work.
Our study represents a step forward by demonstrating the utility of MLLMs for extracting designer-oriented semantic elements vital for inspirational UI design search, using only screenshot images.

\section{Formative Study}
We conducted a formative study to understand how UI designers search for design inspiration and the types of sources, processes, and methods they rely on.
This study focused on identifying the key semantic elements designers consider important during their inspirational search \textbf{(RQ1)}.
Our goal was to gather insights into how designers navigate the search process and the semantics they emphasize, which will later serve as a basis for extracting these semantic elements from an MLLM and establishing design principles to guide the development of an effective search system.

\begin{table}[h]
\vspace{-0.2em}
\caption{Formative Study Participants: Six participants were recruited to capture a broad range of perspectives. YOE denotes years of experience. For students, their number of completed UI design projects is noted.}
\vspace{-0.3em}
\label{tab:formative-participants}
    \begin{tabular}{p{0.04\linewidth}|p{0.37\linewidth}p{0.15\linewidth}p{0.05\linewidth}p{0.13\linewidth}}
    \toprule
    \textbf{ID} & \textbf{Professional Status}      & \textbf{YOE\newline\footnotesize/ \# projects}  & \textbf{Age} & \textbf{Gender} \\ \midrule
    P1          & Product, UI/UX designer          & 8                      & 33           & M            \\ 
    P2          & Product, UI/UX designer          & 8                       & 32           & F          \\ 
    P3          & Product, UI/UX designer          & 5                       & 29           & F          \\ 
    P4          & Senior student      & 2 / 4         & 24           & F          \\ 
    P5          & Junior student      & 2 / 4           & 23           & F          \\ 
    P6          & Sophomore student   & 1 / 3            & 23           & F          \\ 
    \bottomrule
    \end{tabular}
    \Description{This table provides details about the participants in a formative study, which includes six individuals: three experienced UI/UX design professionals and three design major students with varying levels of expertise. The table categorizes participants by ID, professional status, UI design experience, age, and gender. For design students, the number of completed UI design projects is noted. Professional participants have 5+ years of experience, while students have 1+ to 2+ years of experience and 3 to 4 projects completed.}
    \vspace{-0.5em}
\end{table}

\subsection{Participants and Procedure}
To gain a comprehensive understanding of UI inspirational search practices, we recruited six participants, balancing experienced professionals and emerging designers.
This sample size was chosen based on prior research, which indicates that it effectively captures recurring themes and meaningful insights within a focused qualitative study~\cite{guest2006many,morse2000determining}.
Balancing expertise levels ensured a broad range of perspectives, which enriches the depth and applicability of findings~\cite{creswell2017research, cross2004expertise}.
This diversity allowed us to capture general needs from early-career designers to experienced practitioners, providing a more comprehensive understanding of design practices.
The final participants included three expert UI designers and three design major students with UI design experiences, as shown in~\Cref{tab:formative-participants}.

We conducted semi-structured interviews lasting 1 hour each.
The interviews were designed to be conversational and exploratory, allowing participants to share their experiences and insights freely.
We began each interview by asking participants to walk us through their typical design process, paying particular attention to how they search for inspiration and reference materials.
We then investigated their use of existing UI search tools more deeply, encouraging them to share specific examples of successes and frustrations.

All interviews were audio-recorded with participant consent and later transcribed for analysis.
We employed thematic analysis, iteratively coding the transcripts to identify recurring themes, challenges, and desired features.
The authors conducted semantic coding of the interview data, organizing insights into key categories relevant to the design process. 
This process allowed us to distill key insights that informed our understanding of essential semantic elements for UI search and inspiration.

This study received approval from the Institutional Review Board (IRB).
Participants were fully informed of the study’s purpose and procedures, and their consent was obtained prior to participation.
The participants received compensation equivalent to 15 USD for their time and contribution.

\subsection{Key Insights from Interviews}
\subsubsection{UI Design Process and Inspirational Search}
Participants consistently highlighted the importance of the \textit{Double Diamond} design process~\cite{doublediamond}, which involves both the problem definition and the solution exploration phases.
During the process, designers focus on understanding user needs, defining core features, and considering the target audience of the app (P1, 2, 4-6).
One participant (P2), a designer with eight years of experience, remarked:
\begin{quote}
    \textit{``In the discovery phase, we focus on understanding the user and defining the features of the app. Knowing the target audience shapes much of the design process."}
\end{quote}

This reflects the significance of \textit{App Category} and \textit{Target Audience} in the early design stages, indicating the need for tools that allow designers to search for UIs based on these factors.

During the \textit{solution exploration} phase, designers begin wireframing and developing the visual aspects of the interface.
A common challenge noted by multiple participants was finding screens that served similar functional roles while maintaining their design styles.
As P6 explained:
\begin{quote}
    \textit{``I search for screens with a similar function because it helps me figure out the layout and elements I need."}
\end{quote}
Including this, multiple participants emphasize the importance of \textit{Screen Roles} as a key semantic in UI search tools, helping designers align their screens with the intended function within the user flow (P2, 3, 5, 6).

\subsubsection{Sources of Inspiration and their Limitations}
Designers frequently seek inspiration from established platforms to assist them in their work, such as \textit{Pinterest}, \textit{Behance}, and \textit{Dribbble}.
However, most participants were frustrated with the lack of filtering options that account for UI-related elements like app and screen information, as Wu and Xu \etal~\cite{wu2021exploring} highlighted in earlier research.
As P2, one experienced designer commented:
\begin{quote}
    \textit{``It’s hard to find screens that match my app's category or the role of a specific screen in the user flow. Most platforms just give me visually similar designs, but the context is missing."}
\end{quote}
This highlights the necessity of incorporating application and screen level semantics (\eg, \textit{App Category} and \textit{Screen Role}) into UI search tools, enabling designers to locate references that align with their project's functional requirements.
While most participants utilized UI-specific platforms like Mobbin to get inspiration for app flows and screen functionalities, they noted its lack of support for filtering based on visual elements such as \textit{Mood} or \textit{emotional tone} (P1-5). P4 explained:
\begin{quote}
    \textit{``I need to find UIs that match the mood I want for my app—whether it's professional or playful. But finding something that fits that tone is hard."}
\end{quote}
This underscores the need for a search tool that allows designers to filter UIs by \textit{Mood}, a crucial component of the visual design of apps.

\subsubsection{Challenges with Current Search Methods}
A common issue voiced by participants was the inefficiency of existing UI search methods, which rely heavily on simple keyword-based queries (P1-3, 5, 6). P3, with five years of experience, shared their frustration:

\begin{quote}
    \textit{``Current tools only let you search with keywords like `login screen UI,' but they don’t give you the deeper context—like the app's audience or the screen’s role. It takes forever to find the right match."} 
\end{quote}

This highlights the need for a more semantic-driven search system, where designers can search not only by textual information or visual appearance but also by meaningful attributes such as \textit{Target Audience}, \textit{Screen Role}, and \textit{App Category}.

Participants also emphasized the time-consuming nature of manually filtering through search results to find the right inspiration (P1, 4-6). As P6 explained:

\begin{quote}
    \textit{``I often have to go through tons of irrelevant results to find something that fits my project. It’s a huge waste of time."}
\end{quote}
This points to the importance of a UI search system that can automatically narrow down results based on key semantics, reducing the manual effort required to find relevant designs.

\begin{table*}[h]
\caption{Categorized breakdown of key semantic elements in mobile UI design, detailing \textit{Application}, \textit{Screen}, \textit{Composition}, and \textit{Visual Design} levels, along with their descriptions for a semantic-based retrieval.}
\vspace{1em}
\label{tab:information_table}
\centering
\begin{tabular}{c l l}
\toprule
\textbf{Level} & \textbf{Semantic} & \textbf{Description} \\ 
\midrule
\multirow{4}{*}{Application}      & App Category         & The type of application (\eg, social, productivity, or entertainment).         \\ 
                                & App Description      & A brief overview of the app’s purpose, features, and functionality.            \\ 
                                & Similar App          & Examples of apps with similar functionality or design principles.         \\ 
                                & Target User          & The primary audience the app is intended for.         \\ 
\midrule
\multirow{4}{*}{Screen}   & Screen Category      & The type of screens (\eg, login, dashboard, or profile screen).     \\ 
                                & Screen Role          & The specific purpose of the screen, similar to screen summary.       \\ 
                                & Next Screen          & The screen users will navigate to after completing possible actions.    \\ 
                                & Previous Screen      & The screen from which users arrived, offering context or continuity.    \\ 

\midrule
\multirow{3}{*}{Composition}     & UI Elements         & Specific UI elements such as buttons, text fields, or images.         \\ 
                                & Action Items         & Interactive elements like buttons or links that prompt user actions.        \\ 
                                & Layout               & The arrangement of UI elements on the screen.             \\ 
\midrule
\multirow{3}{*}{Visual Design}  & Color Scheme         & The color combinations used across the UI.      \\ 
                                & Color Palette        & The specific colors chosen for UI elements.      \\ 
                                & Mood                 & The emotional tone conveyed by the app’s design.      \\ 
\bottomrule
\end{tabular}
\Description{
This table provides a structured classification of semantic attributes used in UI design, organized into four hierarchical levels: App level, Screen level, Composition, and Visual Design. Under the App level, attributes include App Category, App Description, Similar App, and Target User. The Screen level covers Screen Category, Screen Role, Next Screen, and Previous Screen. Composition attributes are listed as UI Elements, Action Items, and Layout. Lastly, the Visual Design level comprises Color Scheme, Color Palette, and Mood. This table serves as a detailed guide for semantic-based analysis and categorization in UI design, facilitating targeted retrieval and understanding of user interface elements.
}
\vspace{1em}
\end{table*}

\subsection{Key Semantics Derived from the Study}
Through semantic coding, we identified several key semantics that designers consider important when searching for inspiration.
We found several UI semantic elements that were consistently mentioned across participants, and these elements were organized into four main levels: \textit{Application (App)}, \textit{Screen}, \textit{Composition}, and \textit{Visual Design} level.
At the app level, we identified the app category, description, similar apps, and target user information.
The screen level semantics included screen category, role, and navigation context, while the composition level focused on UI elements and action items.
The visual design level encompassed color (scheme and palette) and mood considerations.

To ensure our semantics reflect widespread designer needs, we derived elements mentioned by five or more participants, excluding less frequently mentioned elements like typography, visual style system (\eg, material design, glassmorphism), iconography, and UX writing tone.
Below, we detail each semantic level and its constituent elements, supported by designers' quotes from our interviews.

\subsubsection{Application Level Semantics}
\begin{itemize}
    \item \textbf{App Category and Description}: Designers often reference UIs within a specific app category, such as social media, e-commerce, or productivity, along with descriptions of the app’s purpose and features. This allows them to find designs that align with the functional and aesthetic expectations of the app type.
    \item \textbf{Target User}: The intended user demographic—whether professionals, children, or older adults—plays a significant role in shaping design decisions. Designers expressed a need to search for UIs that cater to specific audiences.
    \begin{quote}
        \textit{``When I design for older adults, I need to find UIs that are accessible and easy to use. But it’s hard to filter for that in most search tools."} (P3)
    \end{quote}
    \item \textbf{Similar App}: Designers often seek inspiration from apps with similar functionality or design principles. Being able to find UIs from apps in the same domain helps ensure their designs align with industry standards and user expectations.
    \begin{quote}
    \textit{``When designing a new app, I always look at top apps in the same category to see how they handle common features and interactions."} (P1)
    \end{quote}
\end{itemize}

\subsubsection{Screen Level Semantics}
\begin{itemize}
    \item \textbf{Screen Category and Role}: Designers often look for UIs based on the functional category and specific role of a screen within the app, such as login screens, dashboards, or profile pages. These semantics are particularly important during wireframing, where the structure and layout are closely tied to the screen’s purpose.
    \item \textbf{Navigation Context}: Participants noted the importance of understanding how a screen fits into the overall user flow, including what screens come before or after it. This helps create cohesive designs that guide users smoothly through the app.
    \begin{quote}
        \textit{``It would be helpful to know what screens come before or after the one I’m designing, so I can better plan the user flow."} (P5)
    \end{quote}
\end{itemize}

\subsubsection{Composition Level Semantics}
\begin{itemize}
    \item \textbf{UI Elements}: Designers frequently search for specific UI elements like buttons, input fields, and navigation bars, which play a key role in the overall usability and functionality of the app. The ability to filter search results based on these components was a widely requested feature.
    \item \textbf{Action Items}: The visibility and emphasis of actionable elements, such as buttons and links, are critical for ensuring user engagement. Designers expressed the need to find UIs that make these elements stand out clearly.
    \begin{quote}
        \textit{“Actionable elements like buttons need to be obvious, and I’m always checking how other apps highlight them.”} (P2)
    \end{quote}
    \item \textbf{Layout}: Designers often reference layouts during wireframing to understand how UI elements are arranged on a screen and how they interact to support functionality and aesthetics. This helps ensure the composition aligns with the intended user flow and design goals.
\end{itemize}
\subsubsection{Visual Design Level Semantics}
\begin{itemize}
\item \textbf{Color}: The color scheme and palette play a crucial role in setting the visual tone of the app. Designers need to find UIs that employ colors effectively to convey the right mood and ensure accessibility.
\begin{quote}
\textit{``Color is so important for creating the right feel. I'm always searching for examples of color schemes that fit the mood I'm going for."} (P5)
\end{quote}
\item \textbf{Mood}: Participants emphasized the importance of finding UIs that convey the right emotional tone or visual mood for their project. The ability to filter UIs by mood helps designers create interfaces that align with the brand’s identity or the app’s goals.
\begin{quote}
\textit{``The mood of the app is key—whether it’s serious and professional or fun and playful. I need UIs that match that mood."} (P4)
\end{quote}
\end{itemize}

By incorporating these semantics, our MLLM-based search system will allow designers to find more relevant and contextually appropriate UI designs, addressing the limitations of current keyword-based tools.
This will enhance the search experience by enabling designers to filter results based on deeper, more meaningful attributes, resulting in a more efficient and targeted inspirational search process.
In addition to the insights gathered from our study, we also incorporated widely accepted semantic elements commonly used in the UI design community.
This comprehensive approach ensures that our system reflects both the nuanced needs of designers, as highlighted in our interviews and the established practices within the industry.
\Cref{tab:information_table} provides a structured summary of the key semantics derived from our study, categorized into app level, screen level, composition, and visual design attributes. These semantics will form the foundation of our approach to semantic-driven UI retrieval, addressing the limitations of existing tools.

\subsection{Deriving Design Principles}
Building on the findings from the formative study, we established three key design principles to guide the development of a semantic-driven UI search system that addresses the challenges and needs highlighted by designers.

\begin{enumerate}
    \item[\textbf{DP1}] \emph{Considering both Functionality and Aesthetics.}
    Existing inspiration platforms often prioritize aesthetics but overlook functional aspects of UI, while UI-specific platforms focus on functionality but lack support for aesthetics like mood and tone. Designers focus on functionality in the early stages and shift to aesthetics in later stages, but current systems fail to support both aspects effectively. This lack of integration of current tools hinders the iterative design process as designers struggle to revisit and refine their work holistically. Supporting both functionality and aesthetics in an integrated manner allows for richer inspiration at each stage, enabling designs that are both effective and visually compelling.\\

\item[\textbf{DP2}] \emph{Flexible Query for Adjusting Priorities.} Our formative study revealed that designers’ search priorities shift across different design stages, focusing more on functional elements in the early stages and on aesthetic aspects in wireframing and visual refinement. This principle emphasizes the need
for a search approach that adapts to shifting priorities, allowing designers to emphasize certain semantics while maintaining a level of consistency in the overall search results.\\

\item[\textbf{DP3}] \emph{Adaptive and Exploratory Search.}
Designers emphasized the iterative nature of their search process and the difficulty of articulating precise search keywords to find relevant references. This highlights the need for iterative query refinement, allowing designers to adjust their searches dynamically. Additionally, surfacing the semantics of search results can help overcome the limitations of keyword-based searches, facilitating a more adaptive and exploratory discovery process during inspirational searches.
\end{enumerate}

These principles address the challenges identified in the formative study, serving as the foundation for our system design. By embedding these principles into the system design, we aim to enhance both the efficiency and creativity of the inspirational search process for UI designers.

\section{Methods}
This section outlines our methodology for extracting semantics from mobile UI screenshots using a multimodal large language model (MLLM). We describe our approach to structuring and extracting key semantics and explain how these semantics are leveraged to build a semantic-based UI search system.
\subsection{Semantic Extraction}
We adopted standardized definitions and techniques for various semantics and structured prompting strategies to extract key semantics using an MLLM.
The following sections detail how we defined semantic structures and implemented the extraction process.

\subsubsection{Defining Categorical Semantics}
We used widely accepted industry-standard sources in mobile UI design to establish consistent definitions for key areas of categorical semantics and align with designers' everyday workflow.
For the \textit{app category}, we utilized the categorical semantics from the Apple App Store\footnote{https://www.apple.com/app-store/}, which provides a comprehensive and standardized classification system for mobile apps.  
For the \textit{screen category}, we adopted the mobile screen categories from Mobbin, a curated UI repository resource frequently used by UI designers, to provide a practical and comprehensive taxonomy.
We also considered using screen topics from Enrico~\cite{enrico} as an alternative but chose topics from Mobbin as it offered finer granularity.
Lastly, for \textit{UI elements}, we categorized UI components (\eg, buttons, text fields, navigation bars) using Mobbin’s classification.
Given the potential for multiple labels to apply to the same screen, semantic outputs were structured in a list format to accommodate overlapping categories.

\subsubsection{Handling Layouts and Visual Design}
For more complex design attributes, such as layout and mood, we curated semantic definitions from existing design literature and filtered them to ensure relevance to mobile interfaces.
For the \textit{layout}, we adopted UI layout guides from online material\footnote{https://devsquad.com/blog/user-interface-layouts}, curating a set of layouts specific to mobile UI and excluding those intended for desktop or other environments.
We selected guideline articles because describing intricate layouts textually is inherently challenging~\cite{UILayoutGeneration}, and these articles offer semantically meaningful textual interpretations of the layout.
For \textit{mood extraction}, we referenced Nielsen Norman Group’s guidelines on moodboard~\cite{tone-of-voice} to provide a robust set of mood keywords and filter those relevant to mobile design.
For \textit{colors}, we adopted a hybrid approach. 
Color is a critical visual design component, but current LLMs often struggle with precise color extraction~\cite{wang2024scientific}.
To address this issue in the context of color schemes, we leveraged LLMs to generate initial color scheme categories.
This approach takes advantage of the extensive knowledge of the LLM of well-known color schemes, allowing it to generate schemes that align with the general design intent.
For the color palette, inspired by a previous study~\cite{wang2023enabling}, which adopts HTML format for LLMs' UI understanding, we presented color information using HTML color names.
We added structure to multi-faceted attributes such as mood and layout by assigning concise keys to each attribute where needed. For example, a screen's mood can be represented by a key-value pair, where the key is the mood, and the value provides a brief explanation of its design implications (\eg, \textit{``playful: A simple and straightforward design facilitating focus on habits.''}). Using keys helps organize attributes for faster lookup, while detailed descriptions support both embedding-based searches and the adaptive and exploratory search principle.
\Cref{tab:semantic_output_format} provides an overview of the structured output formats for all key UI semantics.

\begin{table}[h]
  \caption{Output formats for key UI semantics. The table outlines how semantic attributes are structured into standardized formats for clarity and usability in design tasks.}
  \label{tab:semantic_output_format}
  \begin{tabular}{l|l}
    \toprule
    \textbf{Semantic}  & \textbf{Output Format} \\
    \midrule
    App Category       & List of app categories \\
    App Description    & Brief description of application \\
    Screen Category    & List of screen types \\
    Screen Role        & Brief description of screen purpose \\
    Target User        & Key-value pairs \footnotesize\texttt{(user types: descriptions)} \\
    Similar Apps       & List of app names with descriptions \\
    UI Elements        & Key-value pairs \footnotesize\texttt{(elements: descriptions)} \\
    Layout             & List of layout types with explanations \\
    Action Items       & Key-value pairs \footnotesize\texttt{(UI elements: action)} \\
    Next Screen & Descriptions of potential next screens\\
    Previous Screen & Descriptions of potential previous screens\\
    Mood               & Key-value pairs \footnotesize\texttt{(moods: descriptions)} \\
    Color Scheme       & Key-value pairs \footnotesize\texttt{(schemes: descriptions)} \\
    Color Palette      & Key-value pairs \footnotesize\texttt{(elements: HTML color)} \\
    \bottomrule
  \end{tabular}
  \Description{
  This table overviews the output format used for different semantic categories extracted from UI screens. It lists categories along with their respective output formats like lists, single-paragraph summaries, and key-value pairs. This helps in standardizing the semantic extraction output for better consistency and usability in UI design tasks.
  }
\end{table}

\begin{figure*}[ht!]
    \centering
    \vspace{-1em}
    \includegraphics[width=\linewidth]{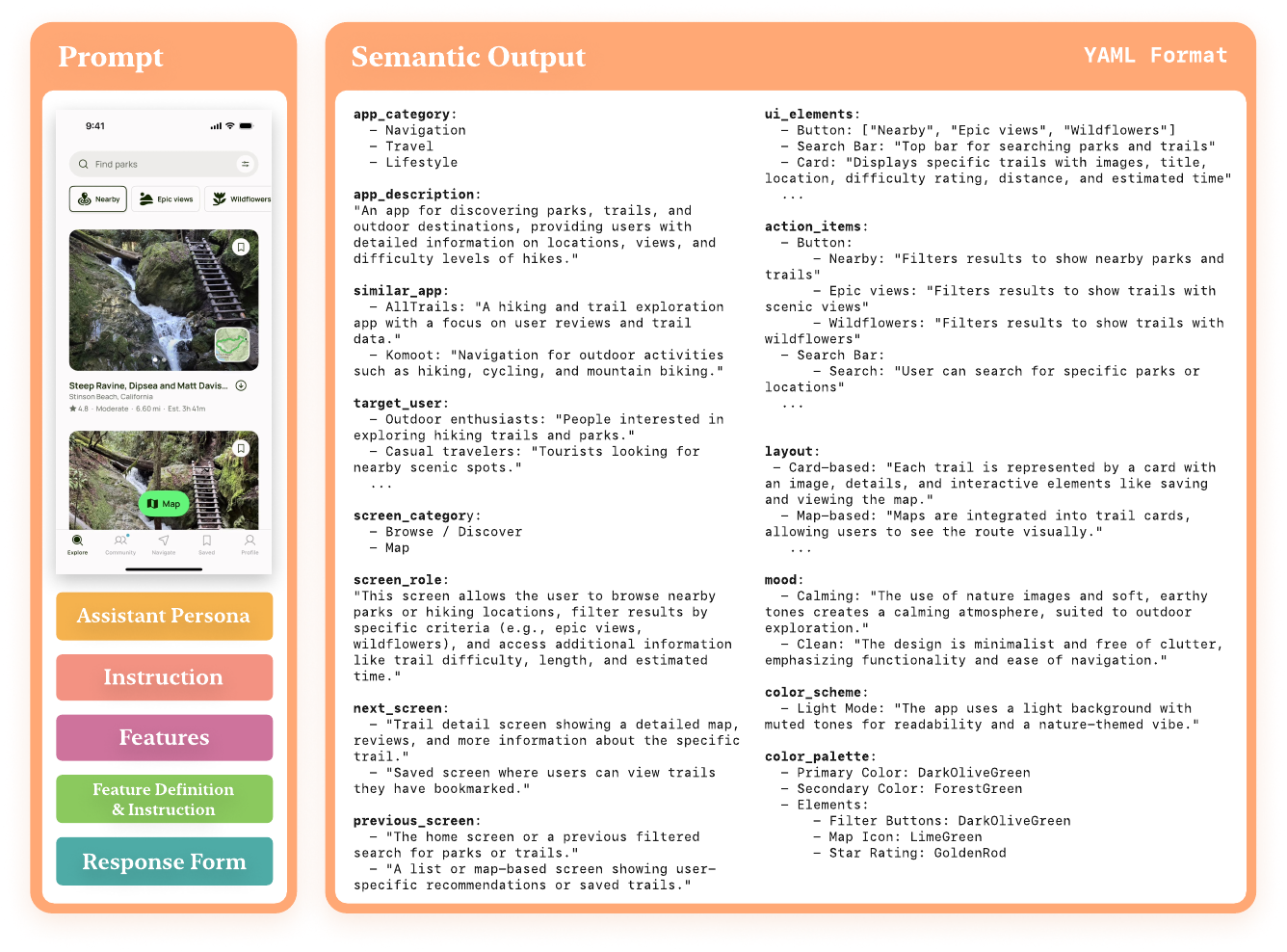}
    \caption{Illustration of prompt and semantic output for mobile UI semantic extraction using multimodal LLM. The left panel displays our prompt concept (the mobile UI screenshot and additional prompts (\ie, assistant persona, instruction, feature list, feature definition \& instruction, and response form), and the right panel presents the structured YAML-formatted output detailing key semantic attributes extracted by GPT-4o.}
    \vspace{-1em}
    \Description{
    Diagram displaying the interaction between user input and the MLLM for extracting UI semantics from a mobile interface. The left side shows a sample mobile UI screenshot and the concept of our prompt. The right side outputs the structured semantic analysis in YAML format, detailing extracted attributes like app category, layout, color scheme, and mood.
    }
    \label{fig:prompt}
\end{figure*}

\subsubsection{Prompting Strategy for MLLM}

To achieve optimal results from the MLLM, we implemented a detailed prompting strategy following established prompt engineering principles that include five key components (\Cref{fig:prompt}): \textit{Assistant Persona}, \textit{Task Instruction}, \textit{Feature List}, \textit{Feature Definition and Instruction}, and \textit{Response Form}.
Each component was encapsulated in an XML format to enhance the efficiency of the prompt execution~\cite{xmlformat}.
The \textit{Assistant Persona} component defined the MLLM as a mobile application design expert specializing in interpreting mobile UI screenshots and identifying detailed characteristics.
This approach is based on the role prompting technique~~\cite{amatriain2024prompt, xu2023expertprompting}, which effectively frames the model’s responses within a specific context and expertise level.
This persona was designed to draw insights from screenshots and provide expert-level analysis tailored to the specific context presented.
The \textit{Task Instruction} component outlined the assistant's task as understanding the screen's content, extracting text, identifying UI elements' roles and positions, and recognizing relationships between them.
We provided clear and specific instructions that are fundamental to effective prompt engineering, guiding the model’s focus and output~\cite{amatriain2024prompt, google_prompt, anthropicClearDirect}.
The \textit{Feature List} component included a comprehensive set of semantic attributes as identified earlier.
The \textit{Feature Definition and Instruction} component provided detailed definitions and instructions for extraction for each feature.
The model was instructed to respond with appropriate values within the predefined categories for categorical semantics.
For example, in the case of UI elements, the model was tasked with identifying each element’s role, while for the next and previous screens, it was guided to align its output with the inferred screen role.
For the color palette, the assistant was prompted to analyze the color distribution first and extract the primary and secondary colors.
It was also instructed to output the colors of major UI components in HTML Color Name format, ensuring clarity and consistency in the response.
This detailed approach to definitions and instructions aligns with the principle of providing comprehensive context to improve the model’s understanding and performance~\cite{fagbohun2024empirical, openaiPrompt, anthropicClearDirect}.
Finally, the \textit{Response Form} component specified the output format.
We chose the output format as YAML, which was selected for its human readability and ability to reduce token counts up to 50\%~\cite{Livshitz_2023}  by eliminating unnecessary characters like braces and reducing blank space indentation of JSON.

\subsubsection{Model and Configuration}
\label{sec:method_model_and_config}
We employed OpenAI GPT-4o~\cite{gpt-4o}, one of the most competitive MLLM models, to extract key semantics from mobile UI images.
During our preliminary testing, we evaluated several publicly accessible MLLMs available in our research period, though the options were just a few.
Among these, GPT-4o demonstrated superior performance in understanding UI concepts and generating consistent semantic descriptions, making it the optimal choice for our research.

To assess the MLLM's ability to understand UI semantics (\textbf{RQ2}), we chose a zero-shot approach, not providing the model with any prior UI examples or semantic labels, for the following reasons.
First, this approach allows us to evaluate MLLMs' inherent capability to grasp UI semantics without the confounding effects of additional training or example data.
Second, fine-tuning or few-shot approaches would require extensive human-labeled datasets, but no well-structured data exists for the semantic attributes we identified.
Moreover, the breadth of semantics cannot be adequately addressed with only a few shots.
A detailed evaluation of the model’s semantic extraction capability in this context is presented in the computational and human evaluation sections.

\begin{figure*}
    \centering
    \includegraphics[width=\linewidth]{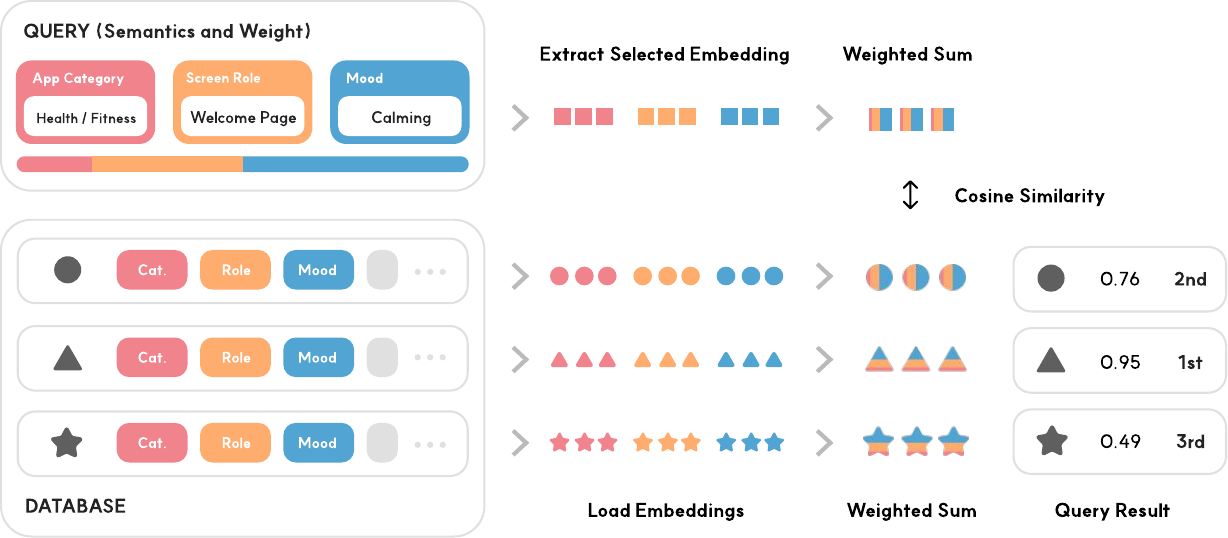}
    \caption{The query and retrieval method illustrates how user-specified semantics and weights are processed to retrieve relevant UI designs from the database.
    The system computes cosine similarity scores between the query and the designs, considering user-adjusted semantic weights to prioritize certain elements.}
    \label{fig:query}
    \Description{
    Flowchart demonstrating the detailed process of querying and retrieving UI designs based on user-specified semantics and weights. The chart starts with the user inputting semantics and weight values. Next, semantic embeddings are extracted for the inputted semantics, which are then used to perform a weighted sum operation with pre-stored embeddings in a database. Cosine similarities between the user’s weighted semantic profile and the database embeddings are calculated, and the UI design with the highest similarity score is selected as the result. This visual representation highlights the computational steps and data flow in the semantic-based retrieval process.
    }
\end{figure*}
\subsection{S\&UI: Semantic-based Retrieval System} %

\subsubsection{Search Query and Retrieval Method Design}
We designed a query system that enables users to input their desired semantics and retrieve the most relevant UI designs from the database.
The search query input consists of user-specified semantics, where users can select the semantics they want to search for and provide a textual description for each semantic, enabling a flexible combination of functional requirements and aesthetic intent (\textbf{DP1}).
To measure the similarity between the user query and the UI designs in the database, we employed the OpenAI \texttt{text-embedding-3-large} model for text embedding.
As shown in~\Cref{fig:query}, we calculated the cosine similarity between the query and semantics in the database.
Guided by the principle of flexible queries for adjusting priorities (\textbf{DP2}), in cases where multiple semantics are involved, users can adjust the weight of each semantic to reflect their importance, and a weighted sum is applied to calculate the overall similarity score.
To construct the search database, we extracted text embeddings for all UI semantics in each screenshot.
For semantics in the key-value format in the YAML output, we represented the semantics by including the key along with the value description.
This approach ensures that the similarity search not only considers the textual description in the value but also reflects the specific semantic represented by the key.
Furthermore, aligned with our principle of supporting functionality aspects of UI (\textbf{DP1}), we developed special queries to address this requirement.
While the regular queries focus on finding UI designs similar to the user-specified semantics, these special queries are designed to retrieve the next or previous screens in the user flow.
This functionality is made possible by the LLM's ability to infer the next and previous screens.
The special queries are implemented by calculating the similarity between the screen roles in the database and the inferred next or previous screen semantics from the query.

\subsubsection{Interactive Search System}

\begin{figure*}
    \vspace{0.5cm}
    \centering
    \includegraphics[width=\linewidth]{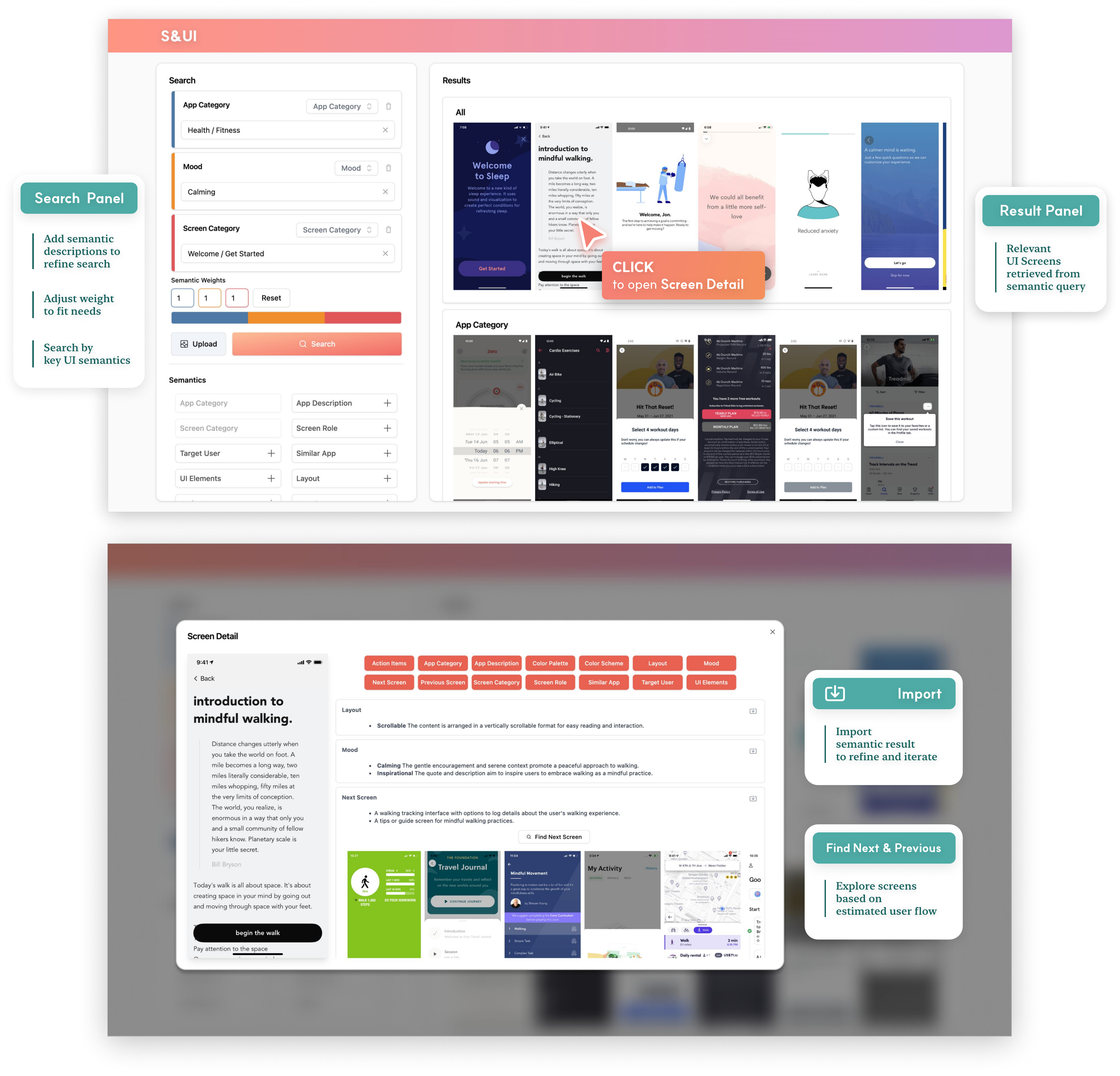}
    \caption{The \systembf system interface. The system enables designers to search UI screens using key UI semantics, such as app category, mood, and screen role. The \textit{Search Panel} allows the addition of semantic descriptions and weight adjustment according to user needs. The \textit{Result Panel} displays relevant screens retrieved based on semantic queries. In the \textit{Screen Detail} Panel, designers can explore detailed screen semantics, iterating and refining their search with the \textit{Import} feature. The \textit{Find Next \& Previous} functionality helps find screens based on estimated user flows, enhancing designers' contextual understanding during the design inspiration process.}
    \Description{
    A detailed view of the S\&UI system interface for semantic-based UI design exploration. The interface is divided into two main panels: a left-side search panel and a right-side results panel. The search panel allows designers to input key semantic attributes such as app category, mood, and screen role and adjust semantic weights to fine-tune search criteria. The results panel displays UI screens that match the search query, sorted by relevance based on the semantic attributes. Clicking on a result opens a detailed screen view, showcasing the comprehensive set of extracted semantics for that design, aiding in-depth analysis and iterative refinement of search criteria. This panel supports navigation through sequential UI designs, enhancing contextual understanding during the inspirational search process.
    }
    \label{fig:interface}
\end{figure*}

We developed an interactive mobile UI search system, \systembf, that embodies our design principles through integrating query formulation, result display, and iterative refinement to facilitate user interaction within our semantic-based retrieval methods, as illustrated in ~\Cref{fig:interface}.

The search panel enables users to define their desired semantics, providing a customizable blend of functional needs and visual preferences (\textbf{DP1}).
Users can freely add or remove semantic attributes, input natural language descriptions for specific design requirements, and utilize dropdowns for predefined categorical semantics.
To adjust search priorities (\textbf{DP2}), the system allows users to assign weights to each semantic, visually represented by a dynamic bar chart above the search button. 
This functionality helps designers fine-tune their queries and prioritize specific UI semantics.
Once the query is submitted, the backend processes the selected semantics and weights to retrieve the most relevant UI designs from the database. The system uses pre-computed text embeddings of the UI semantics for efficient similarity calculation and ranking. 

The results panel presents retrieved UI designs in a structured format, with overall matches considering all provided semantics at the top. Below, individual semantic search results are shown, ranked by similarity scores for each facet.
To support exploratory and adaptive search (\textbf{DP3}), clicking on a result image opens a screen detail panel that displays the extracted semantics for the screen.
This detailed view, as shown in the bottom part of ~\Cref{fig:interface}, allows designers to examine the semantic attributes in-depth, assessing the design's suitability for their project.
To facilitate the iterative nature of design exploration, the import button in the detailed view enables users to incorporate the semantics into subsequent queries, supporting an iterative and exploratory design process.
We also integrated this feature into the search panel to support a query-by-example approach for exploratory searches, enabling users to upload their own UI images.
Upon image upload, the system employs the MLLM to extract semantics, which designers can review and import into the search panel to create structured semantic queries.

\section{Computational Evaluation}
To assess the effectiveness of our proposed method, we conducted a two-step evaluation process.
In the first step, we quantitatively measured UI semantic understanding capability that could be objectively measured using existing UI datasets.
In the second step, detailed in~\Cref{sec:human-evaluation}, we conducted human evaluations with UI designers, focusing on two aspects: (1) assessing the quality of extracted UI semantics and (2) conducting a comparative study where designers conducted searches for design inspiration using our retrieval system \system and an existing tool.
This dual evaluation approach allowed us to understand how the extracted semantics influence the overall search experience.

In this section, to address \textbf{RQ2} on the MLLM's semantic extraction capability, we first conducted computational quantitative evaluations using established metrics and datasets where objective measurement is possible.
This involves evaluations on app and screen category classification, UI element prediction, and screen role understanding.
We utilized relevant existing datasets and established baseline models for measuring capability.
For these evaluations, we utilized the same MLLM (GPT-4o) and configuration as previously mentioned in the~\Cref{sec:method_model_and_config}, to ensure the assessment reflects the model’s inherent semantic understanding capabilities without confounding effects.

\subsection{App and Screen Category Classification}
We utilized the Enrico dataset~\cite{enrico}, which contains 1,460 UI screenshots annotated with screen topics and application metadata.
We mapped the screen topics to our defined screen categories and used the application metadata to evaluate the accuracy of app category extraction.
As a prior study~\cite{spark2023ICML} revealed that foundation models like CLIP~\cite{CLIP} outperform traditional UI-specific models (\eg, Screen2Vec~\cite{Screen2Vec}, Deka \etal~ \cite{RICO}) in retrieving UI screens (based on the app, screen type, and content similarity) tasks, without needing UI structure or other metadata. Based on these findings, we focused our comparison on state-of-the-art vision foundation models, including CLIP models (\texttt{ViT-B/32} (base) and \texttt{ViT-L/14@336px} (most capable)), and GUIClip~\cite{guing}, a \texttt{ViT-B/32} based CLIP model that fine-tuned on UI datasets for retrieving similar UI screens.
We performed zero-shot classification using these models, following the methodology outlined in prior studies~\cite{githubclipprompt, guing, spark2023ICML}.
Specifically, both GPT-4o and the baseline models were provided with identical screen and app category labels from Enrico for classification.
To maintain consistency with baseline models, we prompted simply without other prompt strategies but labels and instructions to output the top-3 categories sorted by confidence.
As there are semantically similar categories which have overlapping functionalities (\eg, \texttt{FORM} and \texttt{LOGIN} for screen category; \texttt{SOCIAL}, \texttt{COMMUNICATION} and \texttt{DATING} for app category), we measured top-1 and top-3 accuracy to test the models' ability to handle semantically overlapping categories.

\begin{table*}
\centering
\caption{Zero-shot classification accuracy comparison of baseline models (CLIP models) and MLLM (GPT-4o) on Screen and App Categories. GPT-4o significantly outperforms baseline models in both top-1 and top-3 accuracy, demonstrating superior semantic understanding of UI screen and app categories.}
\label{tab:zeroshot-accuracy}
\begin{tabular}{lcccccccc}
\toprule
 & \multicolumn{4}{c}{\textbf{Screen Category (20 classes)}} & \multicolumn{4}{c}{\textbf{App Category (31 classes)}} \\
\cmidrule(r){2-5} \cmidrule(l){6-9}
 & \makecell{\footnotesize \textbf{CLIP} \\ \scriptsize \textbf{ViT-B/32}} 
 & \makecell{\footnotesize \textbf{CLIP} \\ \scriptsize \textbf{ViT-L/14@336px}} 
 & \makecell{\footnotesize \textbf{GUIClip} } 
 & \textbf{GPT-4o} 
 & \makecell{\footnotesize \textbf{CLIP} \\ \scriptsize \textbf{ViT-B/32}} 
 & \makecell{\footnotesize \textbf{CLIP} \\ \scriptsize \textbf{ViT-L/14@336px}} 
  & \makecell{\footnotesize \textbf{GUIClip} } 

 & \textbf{GPT-4o} \\
\midrule
Top-1 Accuracy & 26.27 & 29.02 & 36.24 & \textbf{59.21} & 31.01 & 39.95 & 35.65 & \textbf{58.33} \\
Top-3 Accuracy & 55.09 & 55.57 & 58.32 & \textbf{78.83} & 49.91 &  60.57 &  56.70 & \textbf{75.26} \\
\bottomrule
\end{tabular}
\Description{
This table compares the zero-shot classification accuracy of CLIP and GPT-4o models on Screen and App Categories. It provides a detailed comparison across two different CLIP configurations and GPT-4o, showing top-1 and top-3 accuracy percentages in a tabular format. This data highlights the superior performance of GPT-4o in understanding and classifying UI components based on semantic attributes.
}
\end{table*}

\begin{figure*}
    \centering
    \vspace{2pt}
    \includegraphics[width=0.8\linewidth]{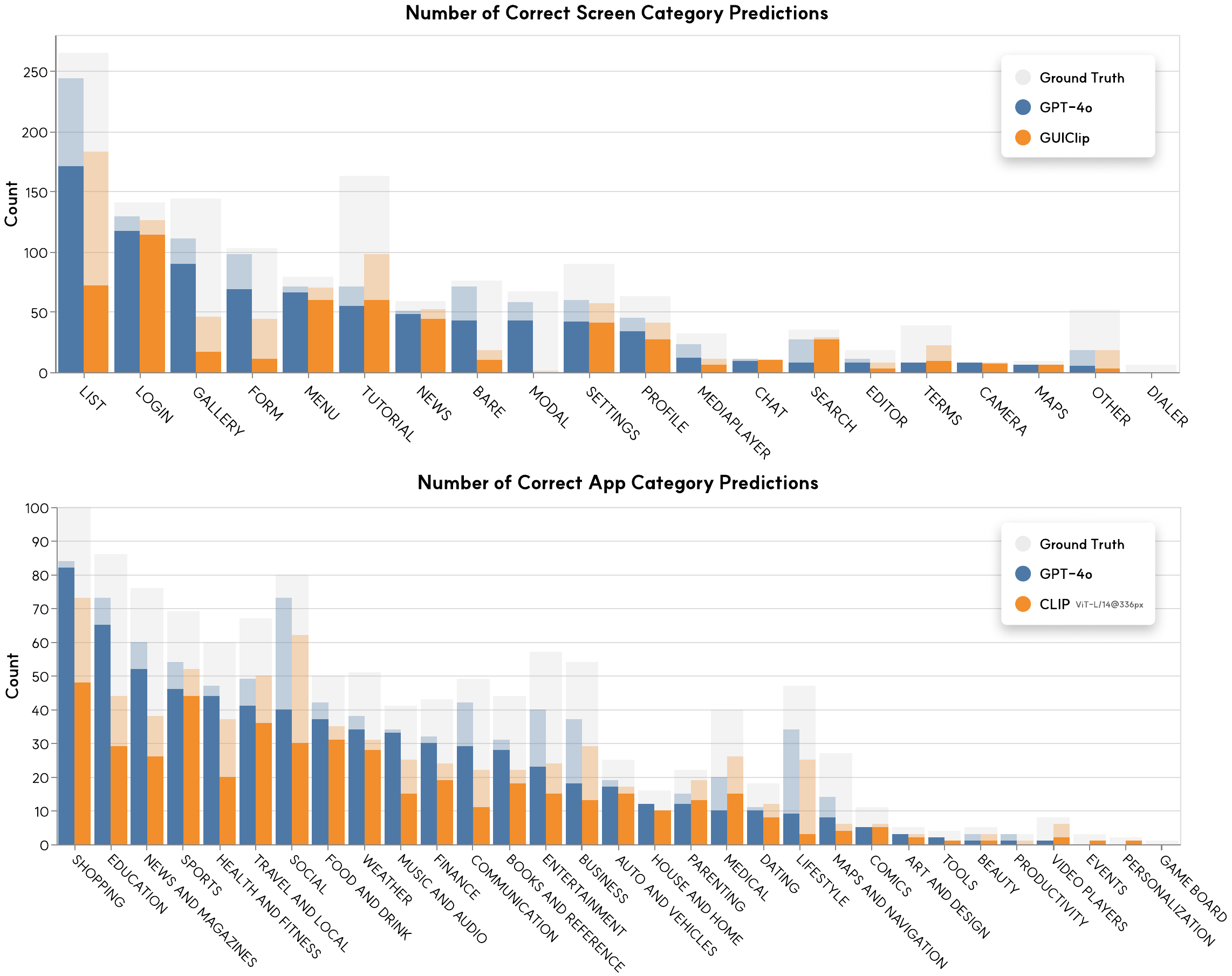}
    \vspace{-5pt}
    \caption{Comparison of Correct Screen and App Category Predictions: The top chart compares GPT-4o and GUIClip across 20 screen categories, while the bottom chart contrasts GPT-4o and CLIP over 31 app categories. Top-1 predictions are highlighted, with top-3 shown semi-transparently. GPT-4o outperforms baseline methods in both cases.}
    \Description{
    A chart that compares correct predictions for Screen and App categories. The top chart evaluates predictions across 20 screen categories, showing that GPT-4o outperforms GUIClip in most categories. The bottom chart compares predictions for 31 app categories, where GPT-4o demonstrates higher accuracy than CLIP. Top-1 predictions are highlighted, with top-3 predictions shown in lighter shading, indicating GPT-4o’s stronger performance in both tasks.
    }
    \label{fig:category_classification}
\end{figure*}

As shown in \Cref{tab:zeroshot-accuracy}, GPT-4o significantly outperformed the baseline models in both screen category and app category classification, achieving a top-1 accuracy of 59.21\% and 58.33\%, respectively, compared to the baselines, which achieved 36.24\% and 39.95\%. 
Moreover, GPT-4o's top-3 accuracy reached 78.83\% for screen categories and 75.26\% for app categories, outperforming the baseline performances of 58.32\% and 60.57\%, respectively.
These results demonstrate that MLLMs can achieve even stronger UI semantic understanding than vision foundation models like CLIP, which have already established significant improvements over traditional UI-specific approaches.
To understand the performance result more deeply, we conducted a detailed category-wise experiment comparing with the most substantial baseline for each task, as shown in~\Cref{fig:category_classification}.
The results reveal that GPT-4o outperforms in most categories.
Interestingly, GUIClip showed higher accuracy in some categories like \texttt{TUTORIAL} and \texttt{SEARCH}; we assume that this is likely due to labeling ambiguities in the Enrico dataset.
\texttt{TUTORIAL} is explained as ``Onboarding screen" can be confused with \texttt{BARE} and \texttt{LOGIN} screens, likewise the ``Search engine functionality" label of \texttt{SEARCH} may not accurately represent basic search interface screens.
Despite these exceptions, the results demonstrate that MLLMs can achieve more substantial UI screen and application understanding than baselines.

\subsection{UI Element Prediction}
We used the CLAY dataset~\cite{CLAY}, which contains annotated UI elements for a subset of the Rico screenshots.
As we focus on inspirational search rather than precise element bounding box prediction, we evaluated the model's ability to identify the presence of different UI element types within a screen.
We provided GPT-4o with the CLAY UI element labels and descriptions, instructing the model to extract only the specified UI element semantics in a zero-shot manner.
We excluded the \texttt{ROOT}, \texttt{BACKGROUND}, and \texttt{CONTAINER} labels from the evaluation as they do not represent typical UI elements, and also excluded \texttt{PICTOGRAM} as the label is confusing with icon and illustration.
The evaluation was conducted on the 1,318 screenshots present in both the Enrico and CLAY datasets.
GPT-4o achieved a weighted average precision of 0.661, recall of 0.703, and $F_{1}$-score of 0.681 across all UI element types.
All metrics are weighted by the number of each label in the dataset.

While these results demonstrate MLLM's ability to identify various UI elements without any task-specific fine-tuning, the absolute performance metrics indicate that there is still room for improvement.
The model's performance is likely limited by several factors, including the zero-shot nature of the task and the inherent ambiguity in some of the UI element descriptions provided by the CLAY dataset. For instance, the distinction between \texttt{TEXT} and \texttt{TEXT\_LABEL} may not be clear based on their textual descriptions alone, which could lead to inconsistencies in predictions.

\begin{figure*}[ht!]
    \centering
    \includegraphics[width=0.8\linewidth]{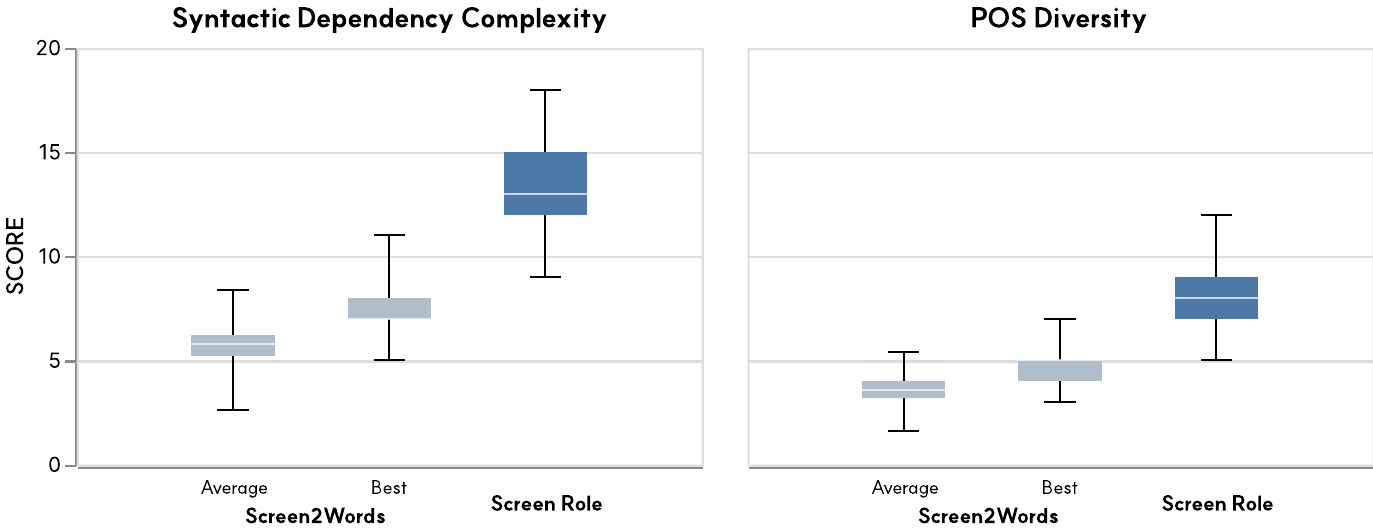}
    \caption{Box plots comparing the syntactic dependency complexity (left) and POS diversity scores of screen descriptions (right) from the Screen2Words dataset (average and best) and GPT-4o. GPT-4o generates more complex and diverse descriptions than the Screen2Words dataset, as indicated by the higher scores in both metrics.}
    \label{fig:screen_role}
    \Description{
    A set of box plots showcasing the comparison between syntactic dependency complexity and Part-of-Speech diversity scores of screen descriptions from the Screen2Words dataset and those generated by GPT-4o. The left plot details the syntactic dependency complexity, while the right plot focuses on Part-of-Speech diversity. Both metrics are higher for GPT-4o, indicating its superior capability in generating more structurally complex and lexically diverse descriptions, which could enhance understanding and classification of screen roles in UI design.
    }
\end{figure*}

\subsection{Screen Role Understanding}
Previous research has explored extracting summaries and descriptions from UI screens.
The Screen2Words dataset~\cite{Screen2Words} provides screen summaries, and studies have shown that LLMs can generate richer results when given screen information in textual form~\cite{wang2023enabling}.

However, the potential of MLLMs in extracting screen descriptions using UI screen images has not been investigated.
Thus, we compare the screen role (description) extracted by GPT-4o with the Screen2Words dataset to evaluate the richness of the generated explanations in a zero-shot manner.
To assess the quality of the screen descriptions generated by GPT-4o, we calculated the syntactic dependency complexity and Part-of-Speech (POS) diversity scores using spacy~\cite{spacy}.
Syntactic dependency complexity measures the intricacy of the syntactic structure, while POS diversity evaluates the variety of lexical types used in the descriptions.
As the Screen2Words dataset contains five descriptions per screen, we compared the average scores and the highest-scoring items from the Screen2Words with the screen role extracted by GPT-4o.

As shown in~\Cref{fig:screen_role}, the results indicate that GPT-4o generates more complex and diverse descriptive text compared to both the average and highest-performing items in the Screen2Words dataset.
The higher scores for syntactic dependency complexity and POS diversity indicate that GPT-4o effectively captures a wider range of syntactic structures and richer lexical repertoires, which leads to more informative and exhaustive screen descriptions.
Our findings suggest that MLLM can generate more detailed and comprehensive screen descriptions than traditional methods by utilizing visual information from UI screen images.

\section{Human Evaluation}
\label{sec:human-evaluation}
To assess the quality and usefulness of the semantics extracted by the MLLM from human perspectives (\textbf{RQ2}) and evaluate the advantages of our semantic-based retrieval system (\textbf{RQ3}), we conducted a two-part human evaluation with UI designers.
To address \textbf{RQ2}, \Cref{sec:quality-assessment} involves UI designers assessing the quality of each extracted semantics. This offers insight into the quality and usefulness of the MLLM-extracted semantics.
\Cref{sec:S&UI-eval} tackles \textbf{RQ3} by employing both our semantic-based search tool and a baseline for realistic inspiration-finding tasks.
Analyzing the ratings and feedback from both systems reveals the advantages and user experience benefits of our approach.

\subsection{Participants}
We recruited 10 participants (4 male, 6 female, denoted E1 - E10) from designer communities and through word-of-mouth, including students, novice designers, and those with professional design experiences,  as detailed in~\Cref{tab:human-evaluator}.
The number of participants and their diverse expertise are considered to align with formative study and practices in computational UI research, where similar studies have included around 10 participants and varying levels of expertise~\cite{wu2024uiclip, wu2024framekit}.
The participants took part in both the 30-minute quality assessment and 1-hour comparative study of retrieval method evaluation, allowing for integrated insights into the semantics and retrieval system.
Participants were compensated at a rate equivalent to USD 22 for the 1.5-hour study session.
The study received approval from the Institutional Review Board (IRB).

\begin{table}
\caption{Human Evaluator Information: Ten participants were recruited to capture a broad range of perspectives. YOE denotes years of experience. For students and novice designers, their number of completed UI design projects is noted.}
\label{tab:human-evaluator}
\begin{tabular}{p{0.04\linewidth}|p{0.37\linewidth}p{0.15\linewidth}p{0.05\linewidth}p{0.13\linewidth}}
    \toprule
    \textbf{ID} & \textbf{Professional Status}      & \textbf{YOE\newline\footnotesize / \# projects}  & \textbf{Age} & \textbf{Gender} \\ \midrule
E1          & Product, UI/UX designer                   & 5          & 32           & M          \\ 
E2          & Product, UI/UX designer                   & 1 / 3   & 25           & F          \\ 
E3          & Product, UI/UX designer                   & 7          & 29           & F          \\ 
E4          & Senior student      & 3 / 5 & 26           & F          \\ 
E5          & Junior student      & 1 / 2  & 26           & M          \\ 
E6          & Junior student      & 2 / 3  & 23           & M          \\ 
E7          & Product, UI/UX designer                   & 12         & 39           & M          \\ 
E8          & Product, UI/UX designer                   & 3 / 6  & 24           & F          \\ 
E9          & Product, UI/UX designer                   & 5          & 27           & F          \\ 
E10         & Product, UI/UX designer                   & 7          & 28           & F          \\ 
\bottomrule
\end{tabular}
\Description{This table summarizes participant information, including ten individuals with diverse design experiences ranging from novice designers and students to professionals. The table includes their professional status, UI design experience (noting years and number of completed projects for students and novices), age, and gender. This selection captures a broad range of perspectives, highlighting varied expertise levels in UI/UX design.}
\vspace{-2em}
\end{table}

\subsection{Quality Assessment of Extracted Semantics}
\label{sec:quality-assessment}
We conducted a quantitative evaluation with UI designers to assess the quality and usefulness of the semantics extracted by GPT-4o.
The evaluation aimed to determine the \textit{Relevance}, \textit{Comprehensiveness}, and \textit{Serendipity} of the extracted semantics at the app, screen, composition, and visual design levels.
For the quality assessment, we randomly selected 1,000 screenshots from the Mobbin and SCapRepo~\cite{guing} datasets, respectively, and extracted semantics using GPT-4o.
Participants freely explored the screenshots and associated semantics, rating at least 15 screenshots per each semantic.
We asked to measure \textit{Relevance} as how accurately the extracted information related to the screen, \textit{Comprehensiveness} as the completeness of the extracted information without missing key details, and \textit{Serendipity} as whether the model uncovered useful information that designers might not have initially considered.
They assessed the 3 criteria of each of the 14 semantics on a 7-point Likert scale, where 1 indicated low, and 7 indicated high.
After the assessment, we conducted semi-structured interviews lasting 5-10 minutes to gather qualitative feedback on the semantics at each level.

\begin{figure*}
    \centering
    \includegraphics[width=\linewidth]{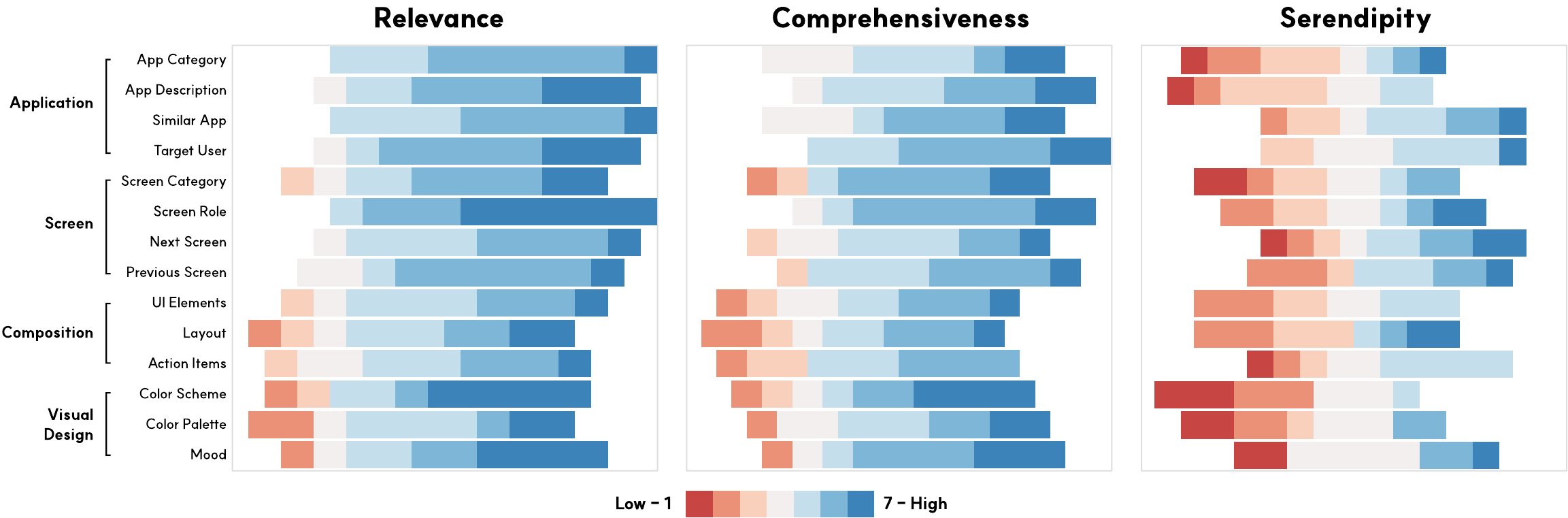}
    \caption{Quality assessment results of extracted semantics by UI designers. The graph shows the rating distribution on a 7-point Likert scale (1-Low to 7-High) for \textit{Relevance}, \textit{Comprehensiveness}, and \textit{Serendipity} across 14 semantic categories grouped into Application, Screen, Composition, and Visual Design levels. Each semantic category is represented by a horizontal stacked bar, with colors ranging from dark red (low ratings) to dark blue (high ratings).}
    \label{fig:semantic-quantitative}
    \Description{
   A horizontal stacked bar chart depicting the quality assessment of semantics extracted by UI designers evaluated on relevance, comprehensiveness, and serendipity across 14 semantic categories such as App, Screen, Composition, and Visual Design. Each category is evaluated on a 7-point Likert scale, represented by colors ranging from dark red (low) to dark blue (high). The chart shows that relevance and comprehensiveness generally received high ratings, indicating strong accuracy and completeness in the extracted semantics, while serendipity had moderate ratings.
    }
\end{figure*}

\subsubsection{Quantitative Assessment of Semantics}
The quantitative results, as shown in~\Cref{fig:semantic-quantitative}, demonstrated that the semantics extracted by GPT-4o were highly relevant, comprehensive, and serendipitous across all levels.
Examining the \textit{relevance} scores, we observe that most semantics (except color palette) achieved mean ratings above 5 on a 7-point Likert scale, indicating a strong overall relevance of the extracted information. 
However, there were some notable variations. The screen role semantic received the highest relevance score (6.5$\pm$0.7), followed by the target user semantic (6.0$\pm$0.94).
This suggests that MLLM excels at capturing the core functionality and intended audience of UI screens.
In contrast, the layout and elements semantics had relatively lower relevance scores, implying potential room for improvement in extracting composition-related information.

Shifting our focus to \textit{comprehensiveness}, we observe a similar pattern, with most semantics receiving mean scores above 5, except elements, layout, and action items.
The screen role and target user semantics again stand out with scores of 5.9$\pm$0.88 and 5.9$\pm$0.74 each, reinforcing GPT-4o's effectiveness in capturing comprehensive information about screen functionality and user characteristics. 
The color palette semantic, however, received a lower comprehensiveness score, suggesting that the extracted color information might be less complete compared to other semantic categories.

Notably, the \textit{serendipity} scores exhibit a different distribution than relevance and comprehensiveness.
Categories like app description and screen description, which focus on well-established information, received lower serendipity scores, indicating that GPT-4o provided highly accurate yet expected results in these areas.
In contrast, semantics related to similar app, target user, and screen flow (next screen and previous screen), which surface more unique and less commonly encountered information, scored higher on serendipity (4.6$\pm$1.58, 4.5$\pm$1.18, 4.6$\pm$2.07, and 4.3$\pm$1.89 respectively).
This suggests that MLLM excelled at uncovering unexpected yet valuable information in cases where designers might not have initially considered such details, enhancing its potential for inspiring novel design ideas.
On the other hand, the color scheme semantic received the lowest serendipity score (2.6$\pm$1.50), suggesting that the extracted color schemes align more closely with designers' expectations and provide fewer unexpected ideas.

Comparing across the different levels, we observed that the app and screen level semantics generally received higher relevance and comprehensiveness scores compared to the composition and visual design levels.
This suggests that MLLM is particularly effective at extracting high-level, functional information about apps and screens.
However, the serendipity scores are more evenly distributed across levels, with the composition level semantics (\eg, action items) showing promise for surfacing novel design ideas.

\subsubsection{Qualitative Interview Insights on Extracted Semantics}

To gain a deeper understanding of designers' perceptions of the extracted semantics, we conducted semi-structured interviews following the quantitative evaluation.
We asked participants for their impressions of each semantic across levels, probing aspects such as accuracy, usefulness, limitations, and potential improvements.
We analyzed the interview responses using thematic analysis, identifying recurring themes and patterns across participants’ feedback for each semantic level.
The interviews provided rich insights into each level's strengths, limitations, and potential use cases of the extracted semantics.

\paragraph{Application Level Semantics}
Participants generally found the app-level semantics (\ie, app category, description, and target user) accurate and informative.
They appreciated the level of detail provided in the app descriptions and target user information, which often went beyond their initial expectations (E1-3, 5, 10).
For example, E10 highlighted a case where the target user description for a banking app included specific details about users' concerns for privacy and security.
However, some participants (E1, 4, 8) mentioned that similar app suggestions were sometimes repetitive or included well-known apps, making it difficult to assess their serendipity.
Participants also noted that while the app category and description information was accurate, it might not always be necessary or useful in their design process (E5, 8).
Several participants were impressed by how the model could extract meaningful application-level information from latent details in single UI screenshots, such as background images or small text elements (E2, 7, 10).

\paragraph{Screen Level Semantics}
Most participants perceived the screen-level semantics, particularly the screen category and role, as highly accurate and valuable. (E1-4, 6, 7, 10)
They found that MLLM was able to effectively identify the main functionality and purpose of each screen, even with limited information (E2, 6, 7).
However, opinions were mixed regarding the usefulness of the previous and next screen predictions.
While some participants found them surprisingly accurate (E3, 6, 9), others felt that they were too generic or lacked the necessary context to be truly helpful in the design process (E2, 7, 10).
Some participants noted that the previous screen predictions were less accurate compared to the next screen predictions (E1, 2, 6), likely due to the inherent ambiguity in inferring prior context from a single screen.
Participants suggested that providing more detailed and contextual information about the screen flow could enhance the value of these semantics.

\paragraph{Composition Level Semantics}
Some participants appreciated the comprehensiveness of the UI element and action item extraction, noting that MLLM could identify elements they might have overlooked (E5, 7).
However, some other participants also pointed out instances where the model misclassified or missed certain elements, such as confusing buttons with other interactive elements or failing to capture the hierarchy and relationships between elements (E3, 9, 10).
The layout semantics were generally perceived as accurate but somewhat generic, with participants noting that many mobile app designs follow similar layout patterns (E3, 10).
Some participants suggested that more specific and nuanced descriptions of layout characteristics could provide additional value.

\paragraph{Visual Design Level Semantics}
Participants had mixed opinions about the usefulness and accuracy of the visual design semantics.
While most found the color palette extraction to be precise (E2-6, 8-10), some participants noticed inconsistencies in identifying primary and secondary colors (E4, 7, 10).
They also mentioned that the color scheme descriptions were often too broad or generic to be informative (E7, 8).
The mood semantics generated more diverse reactions, with some participants finding them accurate and insightful (E3, 9), while others perceived them as too generic or inconsistent (E7, 10).
Participants suggested incorporating additional visual elements, such as imagery and typography, could enhance the richness and specificity of mood descriptions.

\paragraph{Insights across Expertise Levels}
The interviews revealed key differences between novice and expert designers in their evaluations of MLLM-extracted UI semantics. Junior designers (under 3 years of experience; E2, 4-6) expressed satisfaction with app and screen level semantics, appreciating the accurate descriptions of app categories and screen flows, which they found valuable for learning UI design patterns.
In contrast, expert designers (E1, 3, 7-10) acknowledged the accuracy of the semantics but suggested screen flows and mood descriptions need to be more specified. However, they recognized potential uses for app level semantics in competitive analysis and for screen level semantics to enhance communication with developers and product managers.\\

Overall, novice designers benefit from detailed insights to learn UI principles, while experts seek actionable areas for specific tasks. Participants noted that MLLM's semantic extraction could improve their design processes but suggested more fine-grained refinement and contextualization.

\subsection{Human Evaluation of Retrieval Method}
\label{sec:S&UI-eval}
To investigate the advantages and usefulness of our semantic-based retrieval system in supporting UI design tasks (\textbf{RQ3}), we conducted a comparative study where designers used both our system and a baseline system to find inspirational materials for given design scenarios.
The formative study revealed that designers prioritize functionality in the early stages and aesthetics in the later stages of the design process.
To assess whether our system can effectively support both aspects, we developed two types of design tasks: (1) flow tasks for finding inspirational materials related to given user interaction flows and (2) style tasks for discovering screens that are related to the desired mood and tone.
To ensure our evaluation covered representative real-world scenarios, we first generated a pool of 100 UI screen design tasks using an LLM, then converted them into text embeddings, and employed clustering to identify common task patterns.
Through this process, we identified five prevalent application domains and their associated interaction sequences, as shown in~\Cref{tab:tasks}.

\begin{table}
\caption{Flow tasks across different application domains}
\label{tab:tasks}
\begin{tabular}{ll}
\toprule
\textbf{Domain} & \textbf{Flow (Interaction Sequence)}  \\
\midrule
E-commerce & Selecting products and checkout \\
Educational & Viewing courses and taking quizzes \\
Music streaming & Exploring playlists and playing music \\
Travel booking & Selecting and booking hotels or flights \\
Food delivery & Choosing restaurants and placing orders \\
\bottomrule
\end{tabular}
\Description{This table lists interaction flow tasks across various application domains. Each domain is paired with its typical interaction sequence: E-commerce involves product listing, cart management, and checkout; Educational apps include course listing and quizzes; Music streaming features playlist browsing and playback; Travel booking encompasses hotel and flight selection with reservations; Food delivery involves restaurant browsing and order placement.}
\vspace{-1em}
\end{table}

For flow tasks, participants were given a specific application domain and its interaction sequence, then asked to search for screens that could inspire their design of that sequence.
For style tasks, participants chose appropriate tone and manner keywords aligned with their design intentions for each application domain.
Then, they searched for screens conveying these specific tones and stylistic attributes.
For each task, participants were provided with a designated blank Figma frame (or canvas) where they collected and arranged selected UI screens.
For flow tasks, participants organized screens to illustrate the progression of user interactions, similar to a user flow diagram.
For style tasks, participants curated and arranged screens to capture their chosen mood and stylistic direction.

To systematically evaluate both systems across different task types, we employed a Latin square design for task allocation.
Each participant completed four tasks in total: two using \system and the other two using a baseline system.
Each participant performed one flow task and one style task per system. For example, if a participant used \system for a Travel app flow task and a Food delivery app style task, they would then use the baseline system for an E-commerce flow task and an Educational app style task. The sequence of system usage was counterbalanced across participants to minimize learning effects.
The baseline system was a GUIClip~\cite{guing} model-based CLIP retrieval~\cite{clip-retrieval} interface supporting natural language text search queries and image search (both based on GUIClip embedding similarity).
The order of the systems used was counterbalanced across participants to minimize potential bias. 
Each query of both systems returned 15 images, with both systems exploring the same database comprising 10k images from Mobbin.
Descriptive figures about the task allocation~(\Cref{fig:task-allocation}) and baseline system~(\Cref{fig:guing-interface}) are illustrated in the~\Cref{sec:appendix}.

\begin{figure*}
    \centering
    \includegraphics[width=0.7\linewidth]{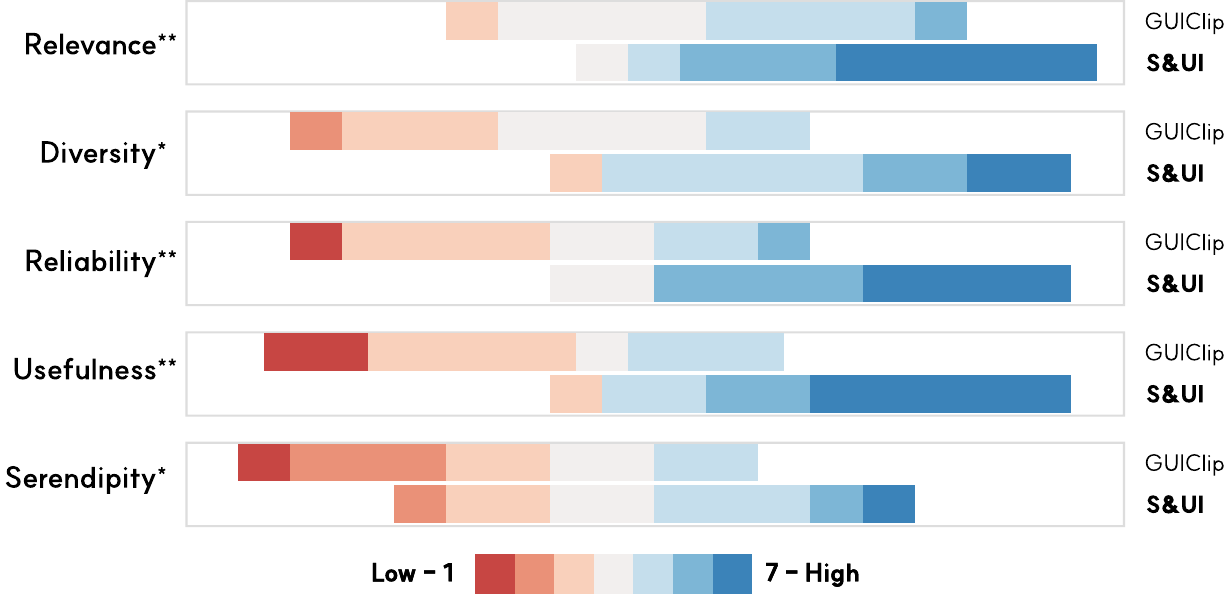}
    \caption{Comparison of user ratings between GUIClip and \system across five metrics: Relevance, Diversity, Reliability, Usefulness, and Serendipity. Ratings are displayed on a 7-point Likert scale (1-Low to 7-High). The stacked bar chart shows the distribution of ratings, with darker shades indicating higher ratings. Statistical significance is denoted by *~$(p<0.05)$ and **~$(p<0.01)$.}
    \label{fig:comparison}
    \Description{
    A stacked bar chart comparing user ratings for two interactive UI design systems, GUIClip and S\&UI, across five metrics: Relevance, Diversity, Reliability, Usefulness, and Serendipity. Ratings are depicted in dark red (low) to dark blue (high) scores on a 7-point Likert scale. The chart shows that S\&UI outperforms GUIClip in all metrics, and all differences between the systems are statistically significant, demonstrating S\&UI’s superior performance in integrating user feedback and enhancing the effectiveness of UI design suggestions.
    }
\end{figure*}

\subsubsection{Quantitative Evaluation of Retrieval Method}
To quantitatively assess the effectiveness of our semantic-based retrieval system \system compared to the baseline GUIClip system, we measured five key metrics: \textit{Relevance}, \textit{Diversity}, \textit{Reliability}, \textit{Usefulness}, and \textit{Serendipity}.
\textit{Relevance} measures how closely the retrieved results align with the given design task and user flow, reflecting the system’s ability to understand and match user intent.
\textit{Diversity} assesses the range of different design ideas presented, ensuring that the system offers a broad spectrum of examples rather than redundant or overly similar designs.
\textit{Usefulness} assesses how each system supports an effective and efficient search experience, focusing on how well they help designers find relevant and diverse design references suited to their needs.
\textit{Reliability} evaluates the system's ability to maintain consistent and appropriate results when designers use semantically similar queries (\eg, ``minimal", ``login screen" vs. ``simple", ``sign-in page") to express the same design intent.
Participants were encouraged to make those multiple queries to assess reliability while searching screens.
Finally, \textit{Serendipity} captures the system’s ability to surface unexpected yet highly relevant results that can inspire designers in novel ways, contributing to creative discovery.

As with the quality assessment study, participants rated each metric using a 7-point Likert scale (1=low, 7=high).
Given the within-subject design and ordinal nature of the ratings, we conducted the Wilcoxon signed-rank test for statistical analysis of the responses.
\Cref{fig:comparison} illustrates the rating distribution for each metric across both systems.
The results show that our \system system consistently outperformed the baseline GUIClip system across all measured dimensions.
The most substantial improvements were observed in relevance, reliability, and usefulness, all of which showed statistical significance at $p<0.01$.
The enhancements in diversity and serendipity still reached statistical significance at $p<0.05$, suggesting that \system offers a broader range of inspirational materials and unexpected discoveries compared to the baseline system.
Overall, these quantitative results strongly support the effectiveness of our semantic-based approach in enhancing the UI design inspiration process.
The consistent outperformance across all metrics demonstrates that \system provides designers with more relevant, reliable, and useful resources while offering diverse and serendipitous discoveries.

\subsubsection{Qualitative Interview Insights on S\&UI}
After completing the tasks with both systems, we conducted semi-structured interviews to gather participants’ experiences with the tools.
We asked questions about the advantages and disadvantages of each system, the quality of the search results, the effectiveness of the semantic-based searches, and how our system compares to existing inspiration platforms.
We conducted a thematic analysis, identifying common themes regarding usability, effectiveness, and potential integration of our system into other design workflows.
The interviews revealed several key themes that highlight the advantages of our semantic-based approach.

\paragraph{Pros and Cons of Semantic-based Search}
Our evaluation demonstrated that the semantic-based system delivers significant advantages that enhance the UI design inspiration process through both its usability and search accuracy.
Participants consistently praised the semantic-based system for its ease of use and ability to find screens that matched their intended queries about design concepts, styles, and flow patterns.
E9 emphasized this, noting how the system consistently delivered exactly what was needed, in contrast to other tools that provided only loosely related results.
They appreciated the wide range of predefined semantic categories and filters, allowing them to narrow their search efficiently.
E5 noted, ``The \system was definitely easier to search with, thanks to the detailed search options."
E7 echoed this sentiment: ``The number of search queries and categories available was impressive, making it simple to find what I needed."
This sophisticated semantic understanding and comprehensive search functionality emerged as a breakthrough feature that enables designers to conduct nuanced, multi-faceted searches that were previously impossible with traditional tools.
Several participants (E2, 3, 7-10) highlighted the system’s weight control feature, emphasizing how it allowed them to refine and focus their searches.
E8 appreciated the weight controls that allowed them to keep their search direction intact while highlighting specific elements they wanted to emphasize.

While \system excels in supporting structured, goal-oriented search tasks, our evaluation revealed specific considerations for early-stage ideation scenarios.
Some participants noted that structured semantic searches might feel too constrained during initial exploration phases when design requirements are still fluid.
As E10 reflected, ``I think the existing tool is more suitable for browsing and getting ideas when you don't have specific requirements in mind",  suggesting opportunities to enhance our system to support more open-ended discovery.
Several participants (E1, 4, 7-9) also expressed interest in adding image similarity features for initial visual inspiration gathering.
These insights suggest opportunities to expand the system’s capabilities to better support undirected and open-ended phases of design ideation.

\paragraph{Support for Iterative Exploration}
Several participants highlighted the value of the iterative search process enabled by \system.
They appreciated how the system allowed them to refine their queries based on previous result screens and their semantics descriptions, helping them clarify their search strategies about design intentions and discover new ideas along the way.
E8 described it as ``narrowing down the search step by step," while E3 noted, ``The ability to iteratively explore and refine my search was really helpful in finding the right inspiration."
This iterative approach was less evident in the baseline system, which often required users to start new searches from scratch when refining their queries.

\paragraph{Potential for Integration into Design Workflows}
Participants saw significant potential for integrating the semantic-based system into their other design workflows than inspirational search.
They identified use cases such as competitive analysis (E3, 10), exploring alternative design solutions (E2, 6), and communicating design ideas with developers (E8, 10).
The system's accurate and relevant results were seen as facilitating better collaboration and alignment within design and development teams.
E2 also highlighted the educational value of the semantic-based system, noting its potential to help novice designers improve their search skills and learn about UI design patterns and best practices.

\paragraph{Search Behaviors Across Expertise}

We found interesting patterns in how designers with varying experience levels utilized \system.
Junior designers (3- years of experience) approached the tool to understand foundational design elements (E2, 5, 6), as E6 noted: ``I search for screens with a similar function because it helps me figure out the layout and elements I need".
Senior designers (5+ years of experience) leveraged more sophisticated search strategies (E3, 7, 9, 10), with E7 stating: ``I combined semantic features together to find exactly what I needed".
This difference in approach was particularly evident when compared to traditional tools, as highlighted by E3, an experienced designer: ``Current tools only let you search with keywords, but they don't give you the deeper context, like the app's audience or the screen's role".
These findings indicate that \system supports both learning and professional workflows through its semantic-based approach.
This adaptability suggests that semantic-based UI search tools can provide meaningful support while accommodating different needs and expertise.

\paragraph{Comparison to Existing Inspiration Platforms}
Participants generally found our system more effective for task-specific UI design compared to both art-focused platforms like Pinterest and Behance, as well as UI curation tools like Mobbin.
Art-focused platforms are considered useful for creative visuals and moodboards, but not as practical for detailed UI and user flow-specific tasks. 
E1 noted, “Pinterest surfaces beautiful images but doesn’t help much when I’m looking for a specific flow or user interface element.”
While Mobbin provided curated UI examples, participants felt it was less flexible in offering deep, customizable search options.
In contrast, our semantic-based system allowed more structured searches through filters like target user, mood, and user flows, which participants found valuable for discovering relevant design references.
E3 highlighted, “I could quickly narrow my search and find exactly the type of screens I needed for my project.”
Overall, while existing platforms served broader inspiration purposes, our system was appreciated for its targeted approach, helping users efficiently find UI designs specific to their design workflows.\\

To sum up, the interview insights from the comparative evaluation highlight the effectiveness and potential of \system in supporting various UI design stages.
Participants consistently preferred the semantic-based system over the baseline, citing its ease of use, accuracy, relevance of results, and support for iterative exploration.
They also recognized its value in enhancing collaboration, communication, and learning within design workflows.

\section{Discussion and Future Work}

Our study leverages MLLM to extract UI semantics from mobile UI images, enhancing design inspiration search and surpassing existing methods.
This section explores MLLM's UI semantic understanding, how we enhanced the design inspiration search process, implications for UI design search tools, and the release of the \system dataset, while discussing limitations and future directions.

\subsection{UI Semantic Understanding with MLLM}

We found that MLLMs can effectively extract meaningful semantics from UI designs.
According to computational evaluation, MLLM shows strength in multiple UI semantic understandings, such as app categories and screen roles, significantly outperforming existing approaches.
Likewise, designers appreciated the model's ability to extract those semantics accurately.
They also appreciated our unique semantics, which had previously been unavailable, like target users, similar apps, screen flows, and mood.
These semantics earned relatively higher serendipity ratings.
Meanwhile, we observed variations in performance across different semantic levels.
In the human evaluation, app and screen level semantics achieved high relevance and comprehensiveness ratings from designers, while composition and visual design level semantics showed few limitations due to their complexity and subjectivity.
The model's extraction capabilities correlate with the concreteness of semantics.
The MLLM performs best when extracting categorical (\eg, app and screen categories) and descriptive semantics (\eg, screen role, app description), but struggles with complex, multiple attributes like layouts, UI elements, and visual design aspects.

Recent advances in vision-language models suggest promising approaches to address these aspects.
For instance, integrating pre-trained vision-language model (VLM) with structured UI data approaches~\cite{gao2024enhancing, lee2023pix2struct, screenAI} to our pipeline could enhance the model's ability to capture hierarchical relationships of UI elements.
Also, incorporating hybrid approaches using both textual model and VLM~\cite{liu2024harnessing, cheng2024seeclick} could better handle specific layout and UI element detection.
Moreover, leveraging emerging direct vision fine-tuning on MLLM~\cite{vision-finetuning}, which has recently become feasible, we could achieve more accurate semantic understandings by fine-tuning using UI-specific datasets for specific semantic attributes while maintaining the flexibility of our approach.
Besides, integrating traditional vision models for color extraction could enhance semantic quality.

These findings indicate that while MLLM can offer significant advantages for extracting meaningful UI semantics, a more detailed approach combining multiple techniques may be needed for comprehensive semantic extraction.
By integrating those approaches into MLLM, future iterations of our research could better address the full spectrum of UI semantic understanding needs, from high-level functional aspects to detailed visual design elements.

\subsection{Enhancing the Design Inspiration Process}
\paragraph{Effectiveness of Semantic-Based Search}
Our user study findings indicate that semantic-based search significantly enhances designers' ability to find relevant inspirational screens.
Designers reported that the system's semantic filters allowed them to perform more granular searches, focusing on specific screen roles, user flows, or target audiences.
For example, a designer looking for "onboarding screens for a fitness app targeting beginners" could retrieve highly relevant examples by searching semantics (screen role: onboarding, app category: health and fitness, target user: fitness beginner), which would be difficult to find using traditional keyword searches.

Combining multiple semantic criteria enabled designers to tailor their search results closely to their project needs.
The weight control feature also allowed participants to emphasize specific aspects of their searches while maintaining consistency in the overall search direction.
This specificity reduced the time spent sifting through irrelevant designs, streamlining the inspiration phase of their workflow.
Participants noted that the system helped them discover designs they might not have found otherwise, enhancing the overall quality of their design exploration.

\paragraph{Explainable UI Search through Semantics}
Providing MLLM's semantic explanations alongside search results increased designers' trust in the system and made the search process more transparent.
Designers appreciated seeing the extracted semantics for each UI screen, which allowed them to understand why specific results were retrieved. This transparency enabled them to decide which designs to consider further.
One participant mentioned, "Seeing the semantic explanation helps me quickly assess if a design is relevant to my project. It feels like the system understands what I'm looking for."
This feedback highlights the importance of explainability in design tools, as it enhances usability and user satisfaction.

\subsection{Implications for UI Design Search Tools}
\paragraph{Integrating Functionality and Aesthetics in Design Tools}
Our findings suggest integrating functional and aesthetic support in design tools aligns well with designers' needs.
Designers often seek inspiration that fulfills specific functional purposes within an app while also being visually appealing.
By providing semantic filters for both aspects, the system helps designers find relevant and aesthetic designs.
For instance, a designer working on a financial app can search for "transaction history screens with a minimalist design," ensuring that the retrieved examples meet both the functional requirements and the desired aesthetic style.
This multi-faceted support facilitates the design process by consolidating search efforts into a single tool.

\paragraph{Supporting Iterative Exploration and Communication}
Providing semantic analysis results for each screen aids designers in iterative exploration.
Designers can refine their searches based on the extracted semantics of previously viewed screens, leading to a more efficient and targeted exploration process.
This iterative approach enables designers to explore broader possibilities while staying aligned with their project goals.
Moreover, the semantic information facilitates better communication between design teams and other stakeholders.
By having a set of semantic descriptors, team members can discuss design elements more effectively.
For example, referring to a "joyful, booking and reserving flow for active people" provides a clear and shared understanding of the design context, reducing misunderstandings and improving collaboration.

\subsection{Public Release of the S\&UI Dataset}

As part of our commitment to advancing research in UI design, we are publicly releasing the \system Dataset used in this study's experiments on GitHub\footnote{https://github.com/spark-damian/S-UI}.
This dataset includes semantic annotations extracted from the UI screenshots used in our evaluations.
By making this dataset available, we aim to provide the research community with a practical resource that can be used to explore new approaches to UI semantic analysis.
This release of \system dataset represents a significant step towards fostering collaboration and innovation in UI design, enabling others to build upon our work and drive new developments in semantic-based UI retrieval, flow analysis, and design comparison.

\subsection{Limitation and Future Directions}

While our studies demonstrated promising results, we acknowledge limitations and areas for future research.
First, our current query design requires manual input of each semantics, which may require additional thought, especially during the early stages of design processes where designers often explore broad ideas rather than specific attributes.
Future work should explore allowing designers to express their needs through free-form queries (\eg, "Show me a minimalist login screen with dark colors"), automatically mapping these to each semantic attribute.
Second, our retrieval method's reliance on weighted embedding similarity becomes less effective as query complexity increases. Alternative approaches, such as semantic combination filtering or category-based matching, could provide more relevant and diverse results.
Also, our current implementation focuses on individual screens, while designers often need to analyze larger patterns across multiple screens and applications. Expanding to multi-screen and cross-app analysis could better support user flow evaluation, including screen transitions, interaction patterns, and design consistency.
Likewise, while our work focuses on mobile UI design, its approaches and techniques could extend to web and desktop applications by diversifying the semantic extraction pipeline to handle more complex layouts and elements.
Another exciting direction is exploring how MLLMs could enable new semantic-aware design synthesis and exploration capabilities in generative design tools.
Additionally, a deeper analysis of how expertise shapes AI-driven design search workflows and use cases could provide valuable insights.
By pursuing these directions, we can work towards more capable, flexible, and intuitive tools that support designers' complex needs and inspire new forms of semantic-aware design exploration.

\section{Conclusion}
We present a novel approach that leverages multimodal large language models (MLLMs) to extract rich semantics from mobile UI images and enables semantic-based inspirational search for UI designers.
Our formative study with professional designers identified key semantic elements crucial for the UI design process, guiding the development of our MLLM-based semantic extraction pipeline and retrieval system.
The evaluation results demonstrate the effectiveness of our extracted semantics and semantic-based search system, outperforming existing methods. 
Our work opens up new opportunities for more intelligent and UI context-aware design tools to accelerate the creative process and empower designers to craft exceptional user experiences.
As the field of UI design evolves and MLLMs advance, we believe our research lays the foundation for next-generation inspirational search systems that can truly understand and support designers' needs and intents.

\begin{acks}
This work was supported by Institute of Information \& communications Technology Planning \& Evaluation (IITP) grant funded by the Korea government (MSIT) [NO.RS-2021-II211343, Artificial Intelligence Graduate School Program (Seoul National University)] and the National Research Foundation of Korea (NRF) grant funded by the Korea government (MSIT) (No. 2023R1A2C200520911).
The ICT at Seoul National University provided research facilities for this study.
We appreciate Sihyeon Lee for his valuable feedback on crafting our prompts and the anonymous reviewers for their thoughtful feedback, which helped improve the quality of this paper.
\end{acks}

\balance
\bibliographystyle{ACM-Reference-Format}
\bibliography{chapters/reference}

\newpage
\appendix
\section{APPENDIX}
\label{sec:appendix}

\begin{wrapfigure}{L}{\textwidth}
    \centering
    \includegraphics[width=\linewidth]{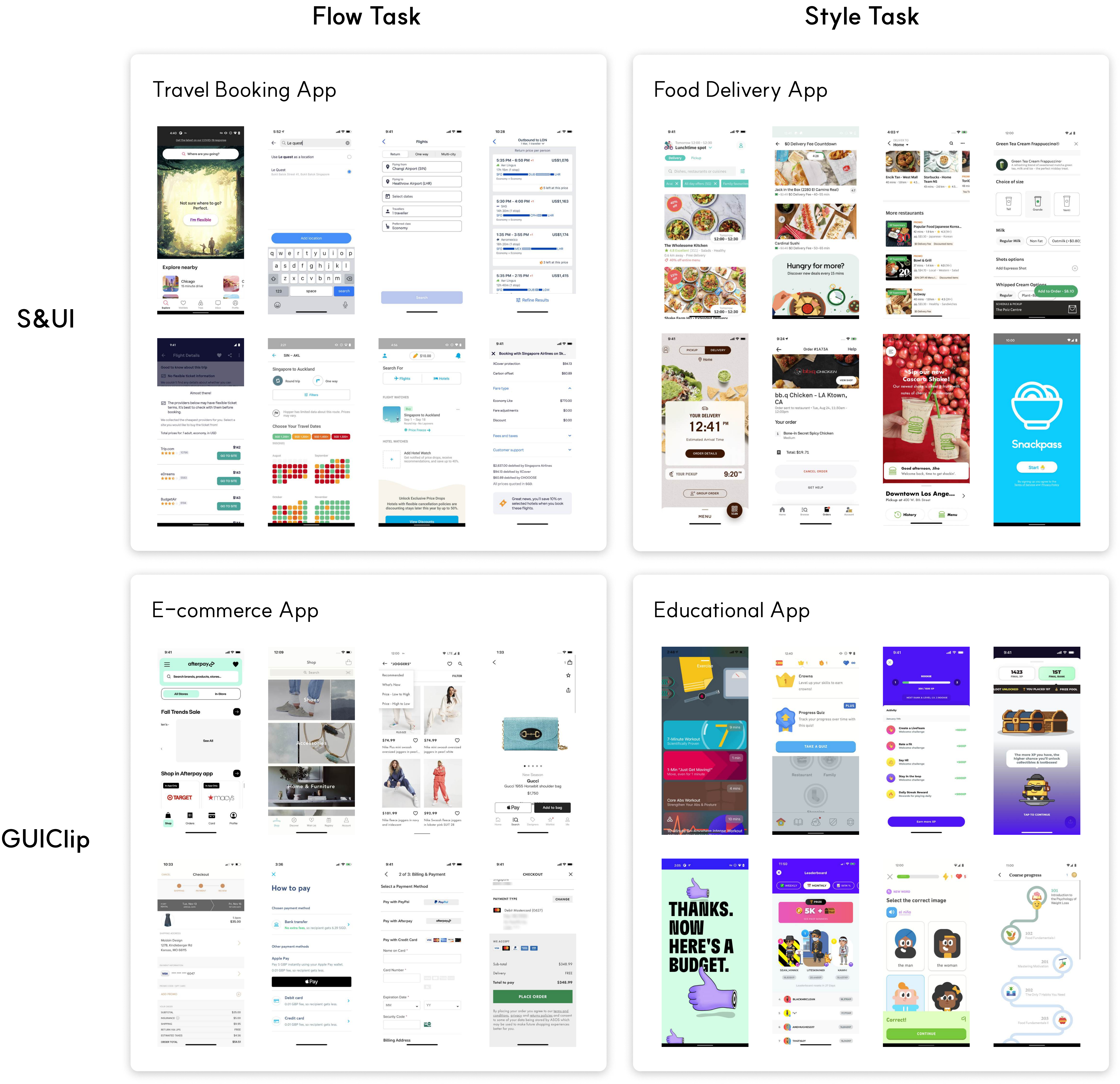}
    \caption{Real examples of results from participant E3 for flow and style tasks for evaluating the \system and GUIClip systems. Flow tasks focused on identifying UI screens to design interaction flows, while style tasks involved discovering UI screens that aligned with specific stylistic intentions. Participants were assigned specific application domains for each task to guide their exploration and arrangement of UI screens.}
    \label{fig:task-allocation}
    \Description{
    The example of results from flow and style tasks for evaluating the S\&UI and GUIClip systems. Flow tasks focus on creating interaction flows for applications like Travel Booking and E-commerce, while style tasks involve identifying stylistically relevant UI screens for applications like Food Delivery and Educational Apps. The examples demonstrate how each system facilitates the exploration and organization of UI screens for specific tasks.
    }
\end{wrapfigure}

\begin{figure*}
    \centering
    \includegraphics[width=0.7\linewidth]{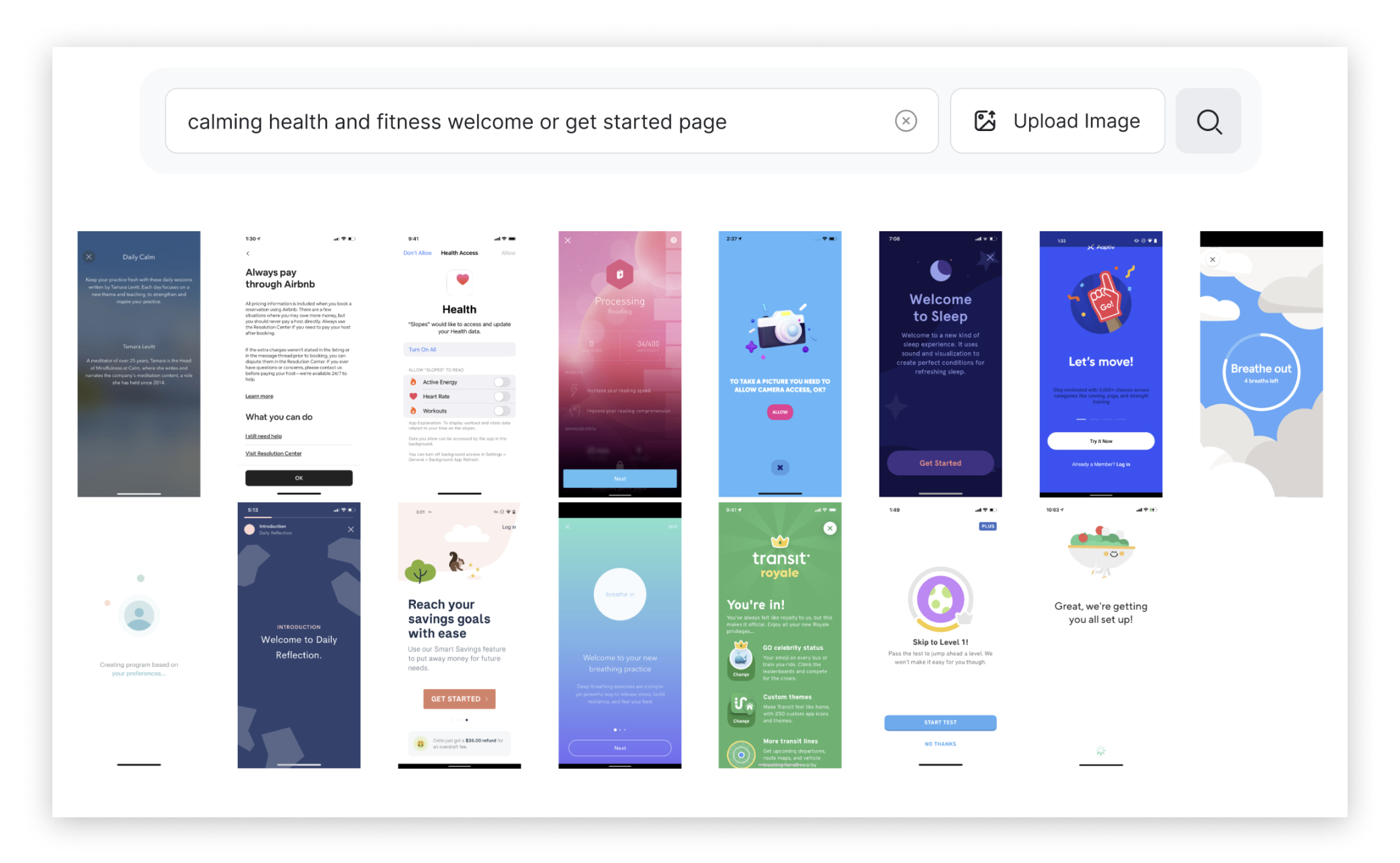}
    \caption{Baseline System Interface. The GUIClip-based retrieval system supports natural language text search queries and image search, utilizing embedding similarity fine-tuned on UI datasets for retrieval. Participants can search for UI screens using descriptive text queries or image queries by uploading or using retrieved result images as new queries.}
    \label{fig:guing-interface}
    \Description{
     The baseline system interface using GUIClip-based retrieval. It supports natural language text search queries and image searches, leveraging embedding similarity fine-tuned on UI datasets. Users can search for UI screens by entering descriptive text queries or uploading images, and they can also use retrieved images as new queries for further exploration.
    }
\end{figure*}

\end{document}